\documentclass[sigplan,screen,10pt]{acmart}
\renewcommand\footnotetextcopyrightpermission[1]{}

\startPage{1}

\setcopyright{rightsretained}

\usepackage{natbib}
\usepackage{url}
\usepackage{amsmath,amsfonts}
\usepackage{algpseudocode}
\usepackage{algorithm,algorithmicx}
\usepackage{graphicx}
\usepackage{textcomp}
\usepackage{tikz}
\usepackage{comment}
\usepackage{xcolor}
\usepackage{subcaption}
\usepackage{booktabs}
\usepackage{array,tabularx,ragged2e}
\usepackage[italic]{mathastext}
\usepackage[export]{adjustbox}
\usepackage{makecell}

\newcolumntype{b}{X}
\newcolumntype{m}{>{\hsize=.55\hsize}X}
\newcolumntype{s}{>{\hsize=.25\hsize}X}

\usepackage{alltt, listings, textcomp, color, verbatim}
\algdef{SE}{Fork}{EndFork}[1]{\textbf{fork} \(\mbox{#1}\) \textbf{do}}{\textbf{join}}%

\lstdefinelanguage{cpp}{
                language=C++,                        
                basicstyle=\ttfamily\tiny,   
                backgroundcolor=\color{black!1},     
                tabsize=2,                           
                showstringspaces=false,              
                showtabs=false,                      
                keywordstyle=\color{blue},           
                identifierstyle=\color{black},       
                emphstyle=\color{black}\bf,          
                commentstyle=\color{gray}\slshape,   
                stringstyle=\color{gray},            
                aboveskip=\baselineskip,             
                xleftmargin=2pt, xrightmargin=2pt, 
                frame=none,                        
                numbers=left,                        
                numberstyle=\tiny,                   
                numbersep=4pt,                      
                captionpos=b,
}
\lstnewenvironment{cpp}{\lstset{style=cpp}}{}

\newcommand{\shared}{\textbf{\_shared\_} }

\usepackage{ifthen,color}
\definecolor{comment}{HTML}{6a9955}
\newcommand{\algcomment}   [1]{{{\color{comment}$\rhd$ #1}}}

\definecolor{darkred}{rgb} {0.5,0.0,0.0}
\definecolor{darkgreen}{rgb} {0.0,0.5,0.0}
\definecolor{darkblue}{rgb} {0.0,0.0,0.5}
\definecolor{orange}{rgb} {1.0,0.65,0.0}

\newcommand{\john}   [1]{{{\color{darkgreen}(john) #1}}}
\newcommand{\osama}   [1]{{{\color{darkred}(osama) #1}}}
\newcommand{\cris}   [1]{{{\color{violet}(cris) #1}}}
\newcommand{\duane}   [1]{{{\color{orange}(duane) #1}}}
\newcommand{\michael}   [1]{{{\color{darkblue}(michael) #1}}}
\newcommand{\final}{0}
\ifthenelse{\equal{\final}{1}}{\renewcommand{\john}[1]{}}{}
\ifthenelse{\equal{\final}{1}}{\renewcommand{\osama}[1]{}}{}
\ifthenelse{\equal{\final}{1}}{\renewcommand{\cris}[1]{}}{}
\ifthenelse{\equal{\final}{1}}{\renewcommand{\duane}[1]{}}{}
\ifthenelse{\equal{\final}{1}}{\renewcommand{\michael}[1]{}}{}

\newcommand{\showcorrections}{0}
\newcommand{\correction}   [1]{{{\color{blue} #1}}}
\ifthenelse{\equal{\showcorrections}{1}}{\renewcommand{\correction}[1]{}}{}

\DeclareFontFamily{OT1}{pzc}{}
\DeclareFontShape{OT1}{pzc}{m}{it}{<-> s * [1.10] pzcmi7t}{}
\DeclareMathAlphabet{\mathpzc}{OT1}{pzc}{m}{it}
\renewcommand*\descriptionlabel[1]{\hspace\leftmargin$#1$}

\begin{document}

\title[Stream-K\@: Work-centric Parallel Decomposition for GEMM on the GPU]
{Stream-K\@: Work-centric Parallel Decomposition for Dense Matrix-Matrix Multiplication on the GPU}


\author[M. Osama]{Muhammad Osama}
\email{mosama@ucdavis.edu}
\orcid{0000-0003-1616-6817}
\affiliation{%
    \institution{University of California, Davis}
    \streetaddress{1 Shields Ave}
    \city{Davis}
    \state{California}
    \country{USA}
    \postcode{95616}
}

\author[D. Merrill]{Duane Merrill}
\email{dumerrill@nvidia.com}
\orcid{0000-0002-4869-9477}
\affiliation{%
    \institution{NVIDIA Corporation}
    \streetaddress{2788 San Tomas Expy}
    \city{Santa Clara}
    \state{California}
    \country{USA}
    \postcode{95051}
}

\author[C. Cecka]{Cris Cecka}
\email{ccecka@nvidia.com}
\orcid{0000-0002-3542-7675}
\affiliation{%
    \institution{NVIDIA Corporation}
    \streetaddress{2788 San Tomas Expy}
    \city{Santa Clara}
    \state{California}
    \country{USA}
    \postcode{95051}
}

\author[M. Garland]{Michael Garland}
\email{mgarland@nvidia.com}
\orcid{0000-0001-6093-7602}
\affiliation{%
    \institution{NVIDIA Corporation}
    \streetaddress{2788 San Tomas Expy}
    \city{Santa Clara}
    \state{California}
    \country{USA}
    \postcode{95051}
}

\author[J. D. Owens]{John D. Owens}
\email{jowens@ucdavis.edu}
\orcid{0000-0001-6582-8237}
\affiliation{%
    \institution{University of California, Davis}
    \streetaddress{1 Shields Ave}
    \city{Davis}
    \state{California}
    \country{USA}
    \postcode{95616}
}


\thanks{Distribution Statement ``A'' (Approved for Public Release, Distribution Unlimited).}

%

\begin{abstract}
    We introduce \emph{Stream-K}, a work-centric parallelization of matrix
    multiplication (GEMM) and related computations in dense linear algebra.
    Whereas contemporary decompositions are primarily tile-based, our method
    operates by partitioning an even share of the aggregate inner loop
    iterations among physical processing elements. This provides a
    near-perfect utilization of computing resources, regardless of how
    efficiently the output tiling for any given problem quantizes across the
    underlying processing elements.

    On GPU processors, our \emph{Stream-K} parallelization of GEMM produces
    a peak speedup of up to 14$\times$ and 6.7$\times$, and an average performance response that is both higher and more consistent
    across 32,824 GEMM problem geometries than state-of-the-art
    math libraries such as CUTLASS and cuBLAS\@. Furthermore, we achieve this performance
    from a \emph{single} tile size configuration per floating-point precision, whereas today's math libraries
    employ complex kernel-selection heuristics to select from a large
    ensemble of kernel variants.
\end{abstract}




\captionsetup{justification=centering}

\maketitle
\pagestyle{plain}

\section{Introduction}

General matrix-matrix product (GEMM), convolution, and other similar 
computations constitute the dominant workloads in many deep learning and 
scientific computing applications. High-performance processors such as GPUs, 
for example, are designed to achieve nearly 100\% of their theoretical peak 
math throughput when computing GEMM.  Doing so, however, requires a work 
decomposition that perfectly occupies the underlying physical cores.  As we 
show, attaining such high levels of processor utilization across a broad 
landscape of problems shapes and sizes can be challenging.

Classically, GEMM implementations block their computation using a 
\emph{data-parallel} tiling of the output matrix, assigning the independent
production of output tiles among concurrent threads (or thread groups)
~\cite{Abdelfattah:2016:KAO,Kerr:2017:CUTLASS,Nath:2010:AIM}.  The work 
per output tile is regular, and tile production tends to dispatch across 
idle physical cores in ``waves''.  The overall workload is well-balanced 
and processor utilization is highest when there are many waves, 
i.e., the number of output tiles greatly oversubscribes the number of cores. 

However, such oversubscription has shrunk considerably as processors have grown 
in size.  An increased core count will require fewer waves to produce a given 
tile count.  Bigger cores will compel larger matrix blocking factors, leading 
to fewer waves of larger tiles.  In general, execution schedules with fewer 
waves are much more likely to suffer from \emph{quantization inefficiency}, 
i.e., the processor underutilization that occurs when the number of output 
tiles is not an even multiple of the number of processor cores.  When the last 
wave is partially full, the unused cores must wait for the remaining threads 
to execute millions (if not billions) of multiply-accumulate (MAC) instructions
before they are able to execute any dependent work.

\begin{figure*}
    \centering
    \begin{subfigure}[t]{\columnwidth}
        \centering
        \includegraphics[width=\columnwidth]{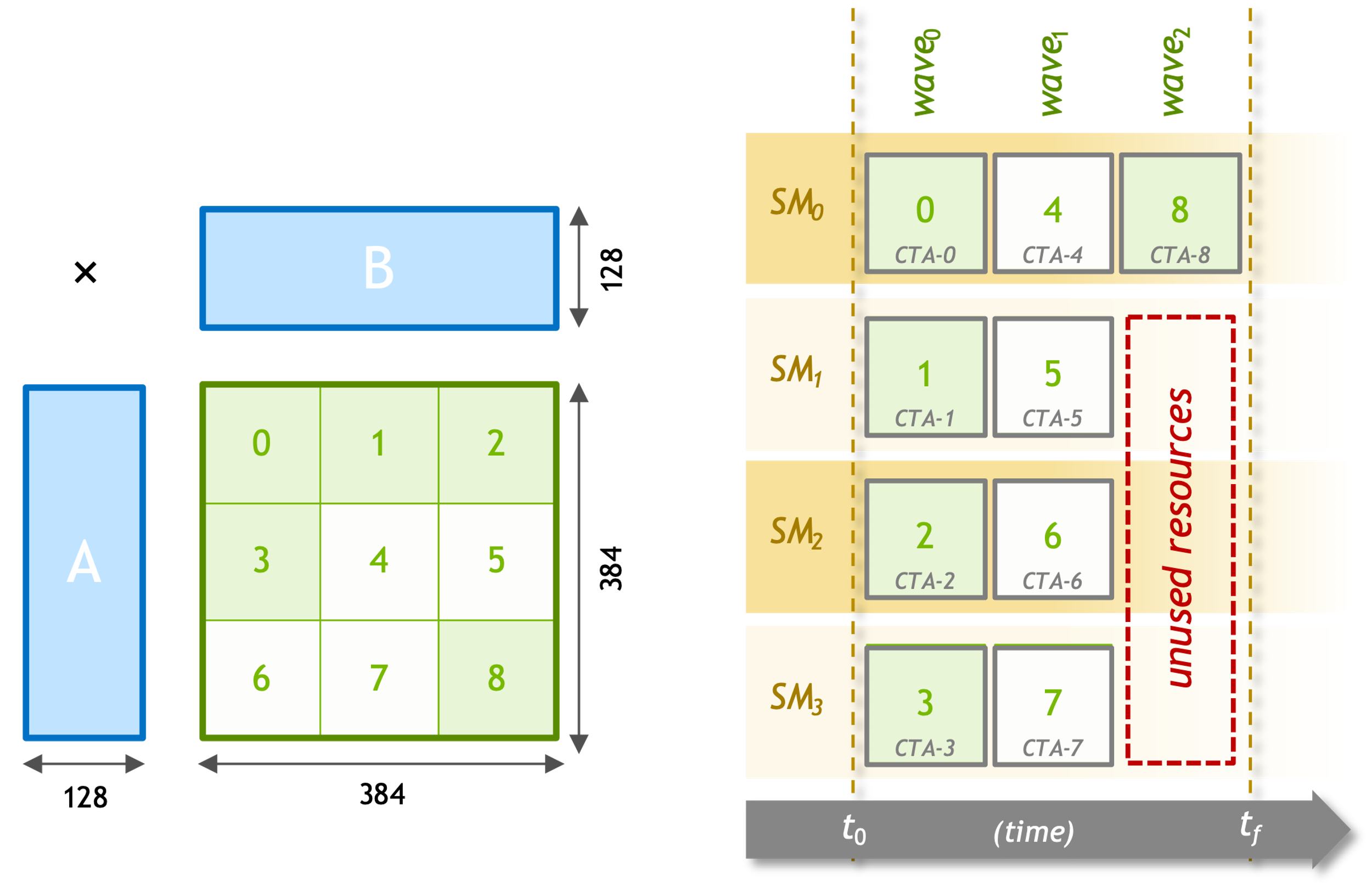}
        \caption{\textit{Data parallel} decomposition with grid size $g$=9~CTAs,\\
          large $128\times128\times128$ CTA work volumes,\\
          and 75\% processor utilization ceiling} \label{fig:data-parallel-efficient}
    \end{subfigure}
    \begin{subfigure}[t]{\columnwidth}
        \centering
        \includegraphics[width=\columnwidth]{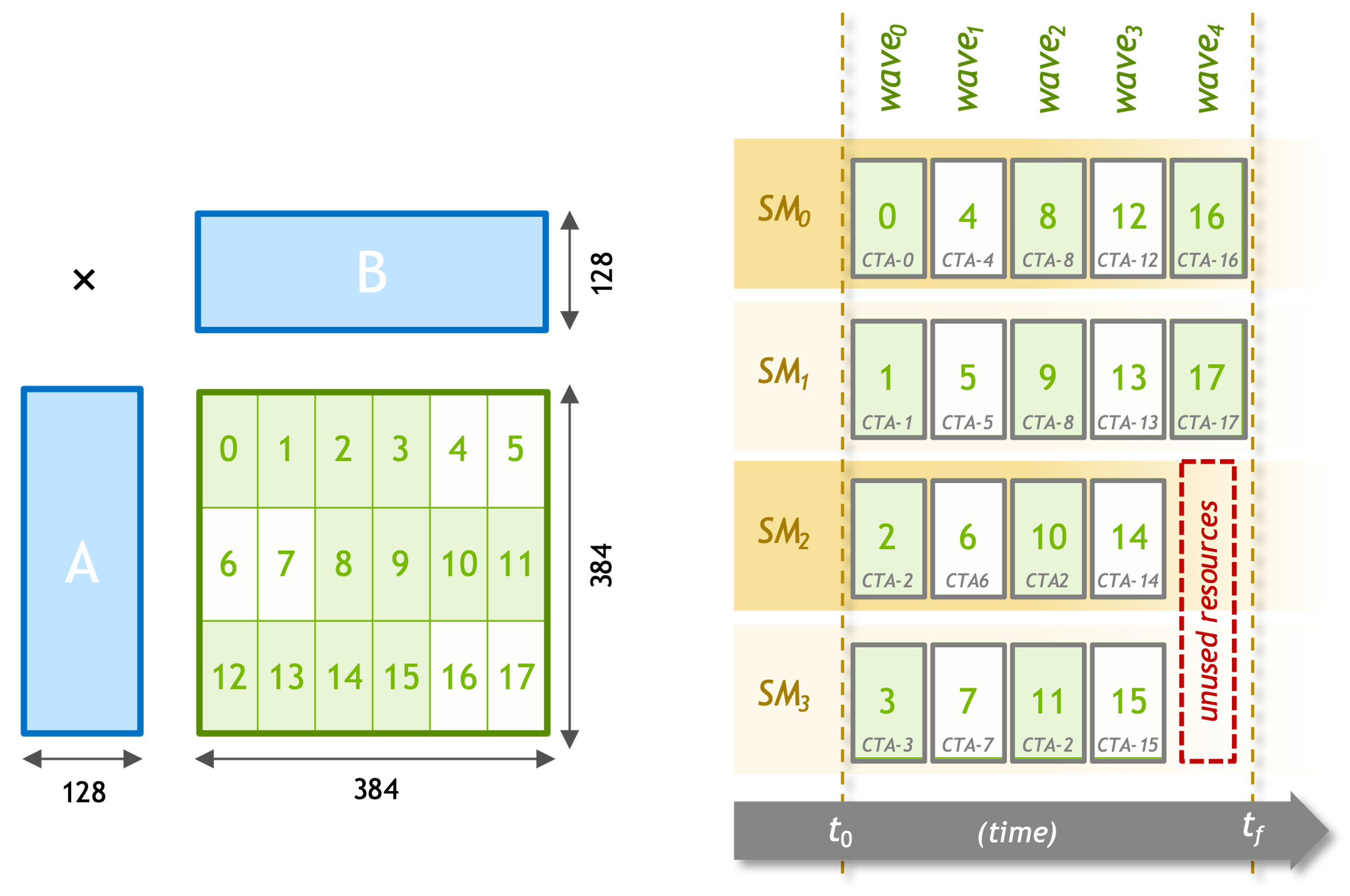}
        \caption{\emph{Data parallel} decomposition with grid size $g$=18~CTAs,\\
          smaller $128\times64\times128$ CTA work volumes,\\
          and 90\% processor utilization ceiling} \label{fig:data-parallel-inefficient}
    \end{subfigure}
    \caption{\emph{Data-parallel} execution schedules for $384\times384\times128$ GEMM across a hypothetical four-SM GPU.} 
    \label{fig:data-parallel}
\end{figure*}

Figure~\ref{fig:data-parallel-efficient} illustrates such a scenario
on a hypothetical GPU with four streaming multiprocessor cores (SMs).
If we block a $384\times384\times128$ GEMM computation into nine
$128\times128$ output tiles, a \emph{data-parallel} decomposition cannot achieve
more than 75\% of the processor's rated throughput.  This theoretical
utilization ceiling can be improved to 90\% by halving the tile size as
shown in Figure~\ref{fig:data-parallel-inefficient}.  However, the
finer-grained blocking factor will be less cache and scratchpad efficient,
and may preclude any practical performance improvement.

\begin{figure*}
  \centering
  \begin{subfigure}[t]{\columnwidth}
    \includegraphics[width=\columnwidth]{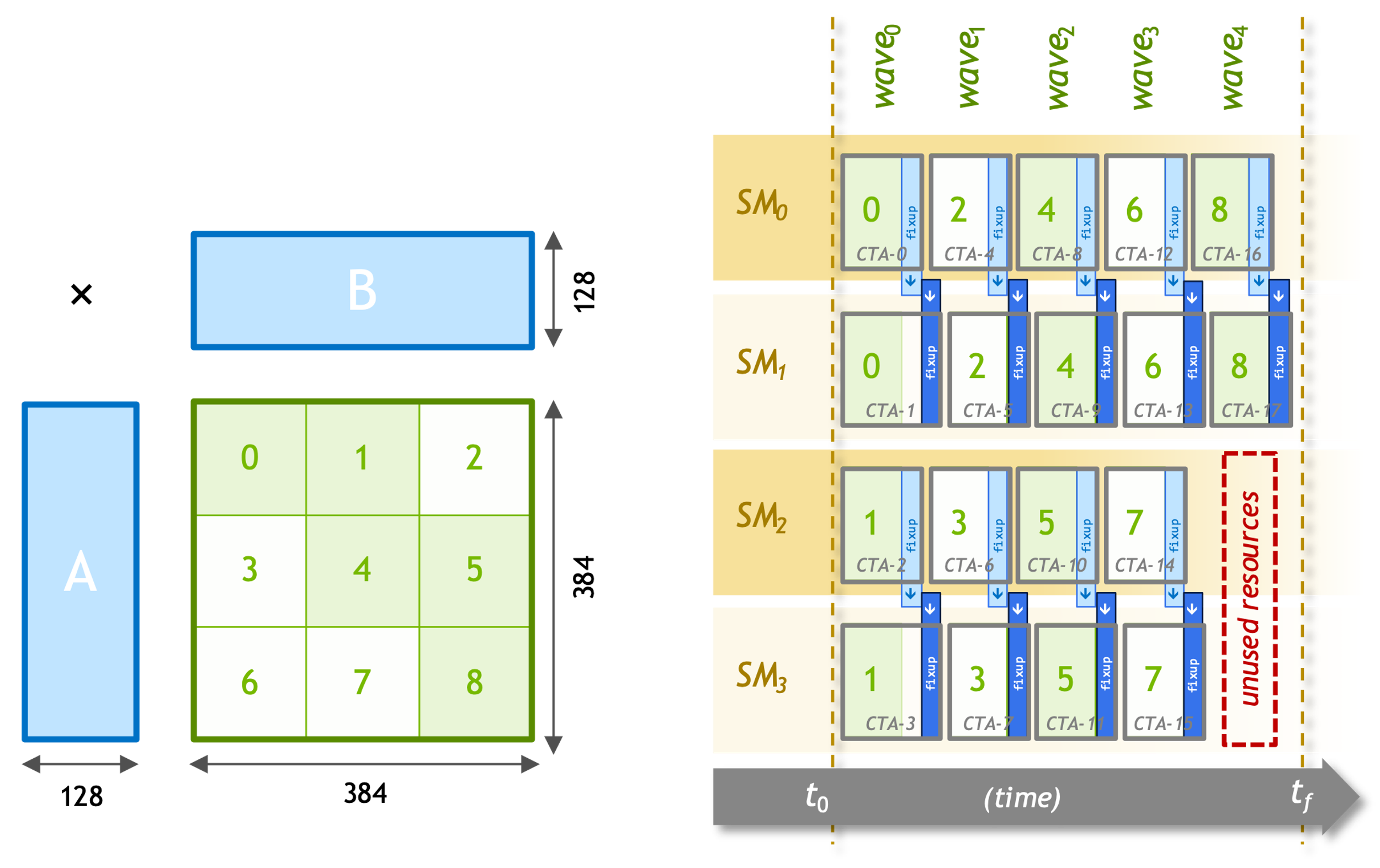}
    \caption{\emph{Fixed-split} decomposition with splitting factor $s$=2,\\
      grid size $g$=18~CTAs, smaller $128\times128\times64$ CTA work volumes,\\
      and 90\% quantization efficiency} \label{fig:fixed_split}
  \end{subfigure}
  \begin{subfigure}[t]{\columnwidth}
    \includegraphics[width=\columnwidth]{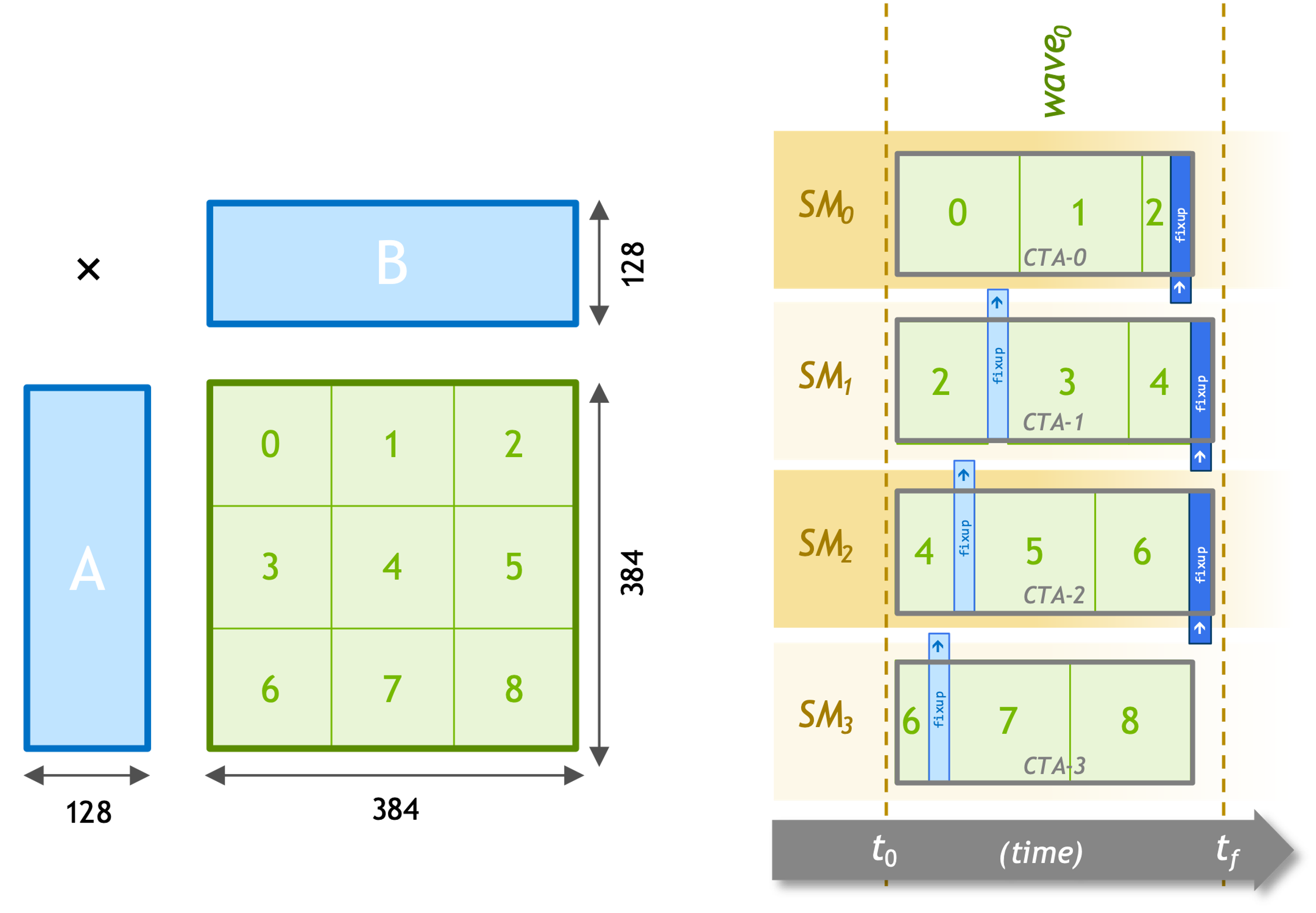}
    \caption{Basic \emph{Stream-K} decomposition with grid size $g$=4~CTAs,\\ 
      larger $128\times128\times288$ CTA work volumes,\\
      and nearly 100\% quantization efficiency} \label{fig:stream_k}
  \end{subfigure}
  \caption{Tile-splitting execution schedules for $384\times384\times128$ GEMM across a hypothetical four-SM GPU.} 
\end{figure*}

Quantization inefficiency is a concern for increasingly wide processors such as GPUs, 
where ALUs-per-core and cores-per-processor both currently number in the 
hundreds. Consequently, many common GEMM-like workloads now exhibit a final, 
partially full wave that comprises a significant fraction of the total computation time.

The current remedy employed by GPU-based math and deep learning libraries is to 
deploy an ensemble of tiling configurations.  When the ideal blocking factor does 
not quantize well, the library chooses among tiling alternatives with smaller 
concurrent work volumes, such as those illustrated in 
Figure~\ref{fig:data-parallel-inefficient} and Figure~\ref{fig:fixed_split}.

Tile-based ensembles, however, present performance and logistical challenges 
for math libraries seeking to deliver the best-achievable performance across diverse 
problem sizes and shapes.  Distributable code size can be problematic for large 
ensembles.  For example, NVIDIA's cuBLAS library~\cite{NVIDIA:2020:CUBLAS} is hundreds 
of megabytes, often providing more than twenty pre-compiled kernel specializations 
per architecture for a given API entry point.  Large ensembles also require 
sophisticated selection heuristics.  In our evaluation, we show these heuristics 
can struggle to consistently identify the optimal configuration for arbitrary 
problems.

Unlike these tile-based methods, our \emph{Stream-K} decomposition always 
distributes an even share (within one) of the aggregate multiply-accumulate loop 
iterations required by the GEMM computation across SMs.  Because the instruction 
workload of a single MAC-loop iteration is far smaller than that of an entire 
output tile, any variance in core workload is practically negligible.  \emph{Stream-K} 
uses the ideal blocking factor regardless of problem shape, has communication 
overheads that scale with processor width (rather than output tiles), and 
compiles to a single kernel.

We use an enormous corpus of 32,824~GEMM shapes and sizes to evaluate \emph{Stream-K}, 
which we implemented within NVIDIA's CUTLASS library~\cite{Kerr:2017:CUTLASS}.  In 
comparison with CUTLASS's \emph{data-parallel} implementation of the same blocking 
factor, \emph{Stream-K} provides a substantially higher performance response across 
our landscape of GEMM problems, demonstrating up to 14$\times$ speedup on NVIDIA A100 
GPUs.

To highlight the practical challenges of ensemble-based solutions, we also 
evaluate NVIDIA's cuBLAS library as well as an oracle-driven 
ensemble of \emph{data-parallel} CUTLASS tilings.  Relative to both ensembles, we 
show that our single-kernel \emph{Stream-K} achieves both (1) higher average 
performance, and (2) higher performance consistency. Versus cuBLAS, \emph{Stream-K} 
demonstrates up to 6.7$\times$ speedup and virtually no instances of slowdown 
for compute-bound problems.

\section{Background}

General Matrix Multiplication (GEMM) is defined as the product $\textbf{C} = \alpha\textbf{AB} + \beta\textbf{C}$ where $\alpha$ and $\beta$ are scalar values and $\textbf{A}$, $\textbf{B}$, and $\textbf{C}$ are matrices. (For simplicity,
we assume $\alpha = 1$, $\beta = 0$ throughout this paper.) We refer to the \emph{shape} of a given GEMM problem by the volumetric extents of its computation. For example, a $m\times n \times k$ GEMM consumes $m \times k$ and $k \times n$ input matrices \textbf{A} and \textbf{B}, respectively, performs $m\times n\times k$ multiply-accumulate operations, and produces an $m \times n$ output matrix \textbf{C}.

GEMM is a performance-critical subroutine in many large-scale engineering and scientific applications. It plays an important role in matrix factorization methods such as LU, QR, and Cholesky decomposition. High-performance modeling and simulation applications in engineering, climate simulation, cosmology, quantum chemistry, and other scientific domains rely on these factorization methods.

Matrix multiplication is also the fundamental building block of modern deep learning (DL) methods. The training of deep neural networks (DNNs) is often performed on massive datasets across large distributed systems~\cite{Mattson:2020:MTB}. Many DL training and inference operations are cast as matrix multiplications. For example, image recognition and computer vision models rely on convolution, which can be implemented directly as the product of filter and image datasets~\cite{Chetlur:2014:CEP}. 
Transformer architectures, which have come to dominate natural language
processing and other applications, are almost entirely limited by the
performance of large matrix products.
%

Early work on GPU matrix-matrix multiplication from Larsen and McAllister framed the computation as a multi-texture multiplication and blending operation~\cite{Larsen:2001:FMM}. The user-programmable shared memory provided by subsequent GPU architectures enabled higher-performing \emph{data parallel} schemes with two levels of blocking (shared memory and registers) with tile sizes informed via extensive micro-benchmarking analysis~\cite{Barrachina:2008:EAT,Nath:2010:AIM,Tan:2011:FID,Tillet:2019:TAI} and auto-tuning~\cite{Cui:2010:ATD,Jiang:2005:ATM,Li:2009:ANA}.

The MAGMA GPU math library was perhaps the first to optimize for diverse GEMM problem shapes~\cite{Kurzak:2012:AGK}. Their solution applied a constrained set of tiling parameters to a templated CUDA C++ code stencil, generating several hundred \emph{data-parallel} variants per API primitive (e.g., \lstinline{hgemm_tt()} for half-precision transpose-transpose GEMM). They evaluated these variants to distill a small ensemble of typically three to five kernels that collectively perform well across a diversity of problem shapes. Kernel selection and dispatch for a given problem was governed by size thresholds expressed via simple handwritten rules.

Subsequent GPU math libraries have employed more sophisticated code-generation 
and kernel-selection components.  For example, the ISAAC project uses machine learning 
techniques to predict an optimal tiling and/or splitting parameterization 
for a given GEMM shape, which can then be instantiated either online or 
offline via a PTX-level code generator~\cite{Tillet:2017:IAA}.

NVIDIA's cuBLAS~\cite{NVIDIA:2020:CUBLAS} library has provided an extended 
\verb|cublasGemmEx| interface that allows the caller to select from among 
24 different GEMM ``algorithms''. Carefully trained heuristics choose between this 
large space of alternatives when using the default interface.  These algorithms 
implement a variety of different \emph{data-parallel} and \emph{fixed-split} variants, 
and it is common for cuBLAS to have assembled each variant into its own 
architecture-specific kernel program for code optimization purposes.  The cross product 
of GEMM API functionality, strategic variants, and microarchitecture has resulted 
in distributions that are increasingly enormous, exceeding hundreds of megabytes of executable code.

Given the fast-paced and rapidly changing nature of contemporary deep learning, recent work has focused on programming models for simplifying the expression and construction high performance kernels that alter or supplement the GEMM computation. The CUTLASS C++ library provides data-movement and multiply-accumulation classes for composing custom GEMM-like computations at all levels of the GPU thread hierarchy~\cite{Kerr:2017:CUTLASS}. Triton~\cite{Tillet:2019:TAI} is a domain-specific language for tensor programming centered on the expression, transformation, and optimization of block/tile concepts. Other domain-specific programming languages such as Halide~\cite{Ragan-Kelley:2013:HAL} and TVM~\cite{Chen:2018:TEO} separate the expression of pointwise operators from that of loop scheduling. Fireiron~\cite{Hagedorn:2020:FAS} further adds data movement constructs into the scheduling grammar.

\section{Existing Work Decomposition Strategies}

Modern processors typically store \textbf{A}, \textbf{B}, and \textbf{C}
in a large, slow, distant memory and have access to a small, fast,
scratchpad or cache memory. A primary goal for any GEMM implementation
is to leverage these local storage resources so that the resulting
implementation is computation-bound.

\subsection{Sequential Cache-Blocked}
The classic cache-blocked formulation of GEMM divides its computational volume 
into blocks and chooses a traversal order that exposes memory locality. 
Algorithm~\ref{alg:sequential} presents a simplified implementation comprising 
six loops. The innermost three loops iterate within the blocking factors BLK\_M, 
BLK\_N, and BLK\_K, while the outermost three iterate across them. If the cache 
can capture one block from each of the three matrices, the resulting data reuse
among those elements will significantly reduce the number of last-level memory 
accesses~\cite{Lam:1991:TCP}.


\begin{algorithm}
    \footnotesize
    \caption{Sequential cache-blocked GEMM\@.}\label{alg:sequential}
    \begin{algorithmic}[1]
    \State \algcomment{tile-processing outer loops}
    \For{mm $\gets$ 0 \textbf{to} m \textbf{step} BLK\_M}
        \For{nn $\gets$ 0 \textbf{to} n \textbf{step} BLK\_N}
            \State \algcomment{zero-initialize output tile}
            \For{mmm $\gets$ mm \textbf{to} (mm + BLK\_M)}
                \For{nnn $\gets$ nn \textbf{to} (nn + BLK\_N)}
                    \State C[mmm,nnn] $\gets$ 0
                \EndFor
            \EndFor
            \State \algcomment{perform the MAC iterations for this tile}
            \For{kk $\gets$ 0 \textbf{to} k \textbf{step} BLK\_K}
                \State \algcomment{MAC iteration (fully unrolled)}
                \For{mmm $\gets$ mm \textbf{to} (mm + BLK\_M)}
                    \For{nnn $\gets$ nn \textbf{to} (nn + BLK\_N)}
                        \For{kkk $\gets$ kk \textbf{to} (kk + BLK\_K)}
                            \State C[mmm,nnn] $\gets$ C[mmm, nnn] +
                            \State \quad (A[mmm,kkk] $\times$ B[kkk,nnn])
                        \EndFor
                    \EndFor
                \EndFor
            \EndFor
        \EndFor
    \EndFor
    \end{algorithmic}
\end{algorithm}

\subsection{Data-parallel}

As shown in Algorithm~\ref{alg:data_parallel}, the \emph{data-parallel} 
GPU formulation of GEMM is decomposed across a grid of parallel thread blocks, 
or \emph{cooperative thread arrays} (CTAs)\footnote{Blocks of GPU threads are 
coscheduled in CTAs, which virtualize the hardware's streaming multiprocessor 
cores (SMs).}. The grid is sized such that each CTA produces its own (BLK\_M 
$\times$ BLK\_N) output tile. 

For exposition, the \lstinline{MacLoop()} subroutine of Algorithm~\ref{alg:macloop}
encapsulates the multiply-accumulate workloads that compute the values of the 
CTA's output tile.  It performs a sequence of \emph{MAC-loop} iterations in the 
accumulation domain, e.g., the \emph{k}-axis for GEMM.  Each \emph{MAC-loop} 
iteration comprises a per-thread volume of (BLK\_M $\times$ BLK\_N $\times$ BLK\_K) 
$/$ CTA\_THREADS MAC operations. As the computation proceeds, fragments of the input 
matrices are staged through the SM's shared memory for local reuse among individual 
threads.

Although this particular presentation of \lstinline{MacLoop()} deploys one thread 
per output tile element, the sophisticated implementations in CUTLASS~\cite{Kerr:2017:CUTLASS} 
and cuBLAS~\cite{Kerr:2017:CUTLASS} will: (1) fully unroll the per-thread MAC-loop 
iteration; (2) implement additional blocking at the warp and/or thread levels; 
and (3) orchestrate a software pipeline of shared memory data movement across 
MAC-loop iterations.

Unfortunately, this classic \emph{data-parallel} decomposition is liable to suffer from 
quantization inefficiency on modern GPUs, as illustrated in Figure~\ref{fig:data-parallel}. 
Although an ensemble of diverse blocking factors may uncover opportunities for greater 
processor utilization, it is unlikely to facilitate perfect quantizations for 
arbitrary problem sizes.  Furthermore, smaller blocking factors have two drawbacks: 
(1)~fewer instructions per MAC-loop iteration for covering the latencies of global and 
shared memory transfers in pipelined implementations; and (2)~a higher proportion of 
memory operations relative to MAC instructions, which may prevent them from being 
computation-bound.



\begin{algorithm}
  \footnotesize
  \caption{\textit{Data-parallel} GPU GEMM\@.}\label{alg:data_parallel}
  \begin{algorithmic}[1]
  \State \shared accum[BLK\_M,BLK\_N]
  \State iters\_per\_tile $\gets \lceil$k/BLK\_K$\rceil$
  \State \algcomment{instantiate one CTA per output tile}
  \Fork{CTA$_{[x]}$ in [ $\lceil$m/BLK\_M$\rceil$ $\times$ $\lceil$n/BLK\_N$\rceil$ ]}
    \State \algcomment{perform the MAC iterations for this tile}
    \State accum $\gets$ MacLoop(x, 0, iters\_per\_tile)
    \State \algcomment{store accumulators to output tile}
    \State StoreTile(C, x, accum)
  \EndFork
  \end{algorithmic}
\end{algorithm}


\begin{algorithm}
    \footnotesize
    \caption{CTA-wide \lstinline{MacLoop()} subroutine for performing a sequence of MAC-loop iterations.}\label{alg:macloop}
    \begin{algorithmic}[1]
    \Procedure{MacLoop}{tile\_idx, iter\_begin, iter\_end}
    \State \shared accum[BLK\_M,BLK\_N]
    \State \shared frag\_a[BLK\_M,BLK\_K]
    \State \shared frag\_b[BLK\_K,BLK\_N]
    \State \algcomment{determine output tile coordinates}
    \State mm $\gets$ BLK\_M $\times$ (tile\_idx / $\lceil$m/BLK\_M$\rceil$)
    \State nn $\gets$ BLK\_N $\times$ (tile\_idx \% $\lceil$m/BLK\_M$\rceil$)
    \State \algcomment{zero-initialize local accumulator storage}
    \State accum $\gets$ {0}
    \State \algcomment{perform the specified range of MAC iters for this tile}
    \For{iter $\gets$ iter\_begin \textbf{to} iter\_end}
        \State kk $\gets$ iter $\times$ BLK\_K
        \State \algcomment{copy global matrix fragments to local storage}
        \State frag\_a $\gets$ LoadFragment(A, mm, kk)
        \State frag\_b $\gets$ LoadFragment(B, kk, nn)
        \Fork{THREAD$_{[mmm,nnn]}$ in [BLK\_M, BLK\_N]}
            \State \algcomment{MAC iteration per thread (fully unrolled)}
            \For{kkk $\gets$ 0 \textbf{to} BLK\_K}
                \State accum[mmm, nnn] $\gets$ accum[mmm,nnn] + 
                \State \quad (frag\_a[mmm,kkk] $\times$ frag\_b[kkk,nnn])
            \EndFor
        \EndFork
    \EndFor
    \State \textbf{return} accum
    \EndProcedure
    \end{algorithmic}
\end{algorithm}

\subsection{Fixed-split}
Alternatively, the granularity of work assigned to each CTA can be reduced via 
parallelization across the accumulation dimension. For a given output tile, the 
associativity of addition allows the iteration domain to be split among multiple
concurrent CTAs, followed by a dependent ``fixup'' step to reduce the partial sums 
computed by each CTA\@. We highlight this \emph{fixed-split} approach in 
Algorithm~\ref{alg:fixed_split}, where each output tile is cooperatively produced
by $s$ CTAs. Notably, it functions identically to the \emph{data-parallel} 
decomposition when the splitting factor $s = 1$.

The \emph{fixed-split} decomposition is also featured in CUTLASS and cuBLAS. The 
splitting factor is implemented as a runtime parameter, allowing a single kernel 
executable to support multiple work volumes while retaining the ideal blocking 
factors for optimal data sharing and latency hiding. However, as illustrated in 
Figure~\ref{fig:fixed_split}, the prospect of achieving a perfect quantization from
a uniform tile-splitting is unlikely.  Furthermore, the extra overheads of 
communication and synchronization scale with both the overall problem size as well 
as the splitting factor.



\begin{algorithm}[!htp]
    \footnotesize
    \caption{\textit{Fixed-split} GPU GEMM with splitting factor $s$.}\label{alg:fixed_split}
    \begin{algorithmic}[1]
    \State \shared accum[BLK\_M,BLK\_N]
    \State iters\_per\_tile $\gets \lceil$k/BLK\_K$\rceil$
    \State iters\_per\_split $\gets \lceil$iters\_per\_tile/$s$$\rceil$
    \State \algcomment{instantiate $s$ CTAs per output tile}
    \Fork{CTA$_{[x,y]}$ in [ $\lceil$m/BLK\_M$\rceil$ $\times$ $\lceil$n/BLK\_N$\rceil$, $s$]}
        \State \algcomment{perform the range of MAC iterations for this split}
        \State iter $\gets$ y $\times$ iters\_per\_split
        \State iter\_end $\gets$ min(iters\_per\_tile, iter + iters\_per\_split)
        \State accum $\gets$ MacLoop(x, iter, iter\_end)
        \State \algcomment{consolidate partial-sums across CTAs}
        \If{y $\neq$ 0}
            \State \algcomment{store accumulators to temporary global storage}
            \State StorePartials(partials[x,y], accum)
            \State Signal(flags[x,y])
        \Else
            \State \algcomment{accumulate partial sums from other CTAs contributing to this tile}
            \For{cta $\gets$ 1 \textbf{to} $s$}
                \State Wait(flags[x,cta])
                \State accum $\gets$ accum + LoadPartials(partials[x,cta])
            \EndFor
            \State \algcomment{store accumulators to output tile}
            \State StoreTile(C, tile\_id, accum)
        \EndIf
    \EndFork
    \end{algorithmic}
\end{algorithm}

\section{Our \emph{Stream-K} Decomposition}
Our \emph{Stream-K} decomposition is a tile-splitting parallelization in which 
the splitting seams are completely dissociated from the tiling structure 
itself. Although we employ familiar blocking and tiling strategies for data 
reuse, we instead quantize the GEMM computation into MAC-loop iterations, 
i.e., small volumes of CTA-wide BLK\_M $\times$ BLK\_N $\times$ BLK\_K work. As 
presented in Algorithm~\ref{alg:streamk}, \emph{Stream-K} evenly partitions the GEMM's 
aggregate workload of MAC-loop iterations across a constant-sized grid of $g$ 
CTAs. Each CTA's range of MAC-loop iterations is mapped contiguously into the 
$m \rightarrow n \rightarrow k$ linearization of the GEMM shape, crossing 
output-tile boundaries as it may.

\begin{algorithm}
    \footnotesize
    \caption{Basic \textit{Stream-K} GPU GEMM with grid size $g$.}\label{alg:streamk}
    \begin{algorithmic}[1]
    \State \shared accum[BLK\_M,BLK\_N]
    \State iters\_per\_tile $\gets \lceil$k/BLK\_K$\rceil$
    \State total\_iters $\gets \lceil$m/BLK\_M$\rceil \times \lceil$n/ BLK\_N$\rceil$ $\times$ iters\_per\_tile
    \State iters\_per\_cta $\gets \lceil$total\_iters / g$\rceil$
    \State \algcomment{instantiate g CTAs}
    \Fork{CTA$_{[x]}$ \textbf{in} [g]}
        \State iter $\gets$ x $\times$ iters\_per\_cta
        \State iter\_end $\gets$ iter + iters\_per\_cta
        \State \algcomment{iteration-processing outer loop}
        \While{iter $<$ iter\_end}
            \State tile\_idx $\gets$ iter / iters\_per\_tile
            \State tile\_iter $\gets$ tile\_idx $\times$ iters\_per\_tile
            \State tile\_iter\_end $\gets$ tile\_iter + iters\_per\_tile
            \State \algcomment{perform the range of MAC iterations for this tile}
            \State local\_iter $\gets$ iter - tile\_iter
            \State local\_iter\_end $\gets$
            \State \quad min(iter\_end, tile\_iter\_end) - tile\_iter
            \State accum $\gets$
            \State \quad MacLoop(tile\_id, local\_iter, local\_iter\_end)
            \State \algcomment{consolidate partial-sums across CTAs}
            \State tile\_started $\gets$ iter = tile\_iter
            \State tile\_ended $\gets$ (iter\_end $\geq$ tile\_iter\_end)
            \If{$\neg$tile\_started}
                \State \algcomment{store accum to temporary global storage}
                \State StorePartials(partials[x], accum)
                \State Signal(flags[x])
            \Else
                \State \algcomment{store accumulators to output tile}
                \If{$\neg$tile\_ended}
                \State \algcomment{accumulate partial sums from other CTA contributing to this tile}
                \State cta\_end $\gets$ tile\_iter\_end / iters\_per\_tile
                \For{cta $\leftarrow$ (x+1) in cta\_end}
                    \State Wait(flags[cta])
                    \State accum $\gets$ accum
                    \State \quad + LoadPartials(partials[cta])
                \EndFor
                \EndIf
                \State StoreTile(C, tile\_id, accum)
            \EndIf
            \State iter $\gets$ tile\_iter\_end
        \EndWhile
    \EndFork
    \end{algorithmic}
\end{algorithm}

Should a given CTA's starting and/or ending iterations not coincide with
tile boundaries (as is expected to be the common case), it must
consolidate its partial results with those of the other CTA(s) also
covering that tile. In this basic implementation, each output tile in
\textbf{C} is written by the CTA that performed that tile's $k=0$
MAC-loop iteration. Before it can do so, however, it must accumulate any
partial sums shared from other CTAs in temporary global storage.
Notably, \emph{Stream-K'}s communication, synchronization, and global
storage overheads are independent of problem size, scaling instead with
the number of CTAs $g$.

A secondary benefit of \emph{Stream-K} is that
synchronization-waiting~is likely negligible when the number of output
tiles is greater than the number of CTAs. In this regime, each output
tile is covered by at most two CTAs, and the tile-processing skew
ensures that the accumulating CTA will not need its peer contributions
until well after those collaborators have finished producing them.

Continuing our earlier example, Figure~\ref{fig:stream_k} illustrates the basic
\emph{Stream-K} execution schedule of the $384\times384\times128$ GEMM problem on a
hypothetical four-SM GPU\@. To fully occupy the GPU, we launch $g=4$
CTAs. Assuming BLK\_M~$=128$, BLK\_N~$=128$, and BLK\_K~$=4$, each CTA is tasked
with a $128\times128\times288$ work volume comprising 72~MAC-loop iterations. This
results in a 100\% quantization efficiency, as all four SMs will execute
the same number of MAC instructions.

Additionally, the work volume of a single MAC-loop iteration is 32$\times$
smaller than that of an entire output tile. Consequently, a 32-way
\emph{fixed-split} decomposition would also provide a 100\% quantization
efficiency, but at the expense of an 8$\times$ larger ``fixup'' overhead.
Furthermore, \emph{Stream-K} is better able to hide the latency of inter-CTA 
synchronization due to the temporal skew between writers and readers when 
sharing partial sums.

\emph{Stream-K} also generalizes to both \emph{fixed-split}
and \emph{data-parallel} decompositions. When the grid size $g$ is
an even multiple of the number of output tiles, \emph{Stream-K}
functions exactly as the \emph{fixed-split} decomposition. Similarly,
when $g$ equals the number of output tiles, \emph{Stream-K} behaves
identically to the \emph{data-parallel} decomposition. We take advantage
of this generalization to create an optimized hybridization of the \emph{Stream-K}
decomposition in following section (\ref{sec:data-parallel-hybridization}).

\section{Implementation Details}
\label{sec:practical-usage}   

\begin{figure*}[hbt!]
    \centering
    \begin{subfigure}[t]{0.15\textwidth}
        \includegraphics[width=\linewidth]{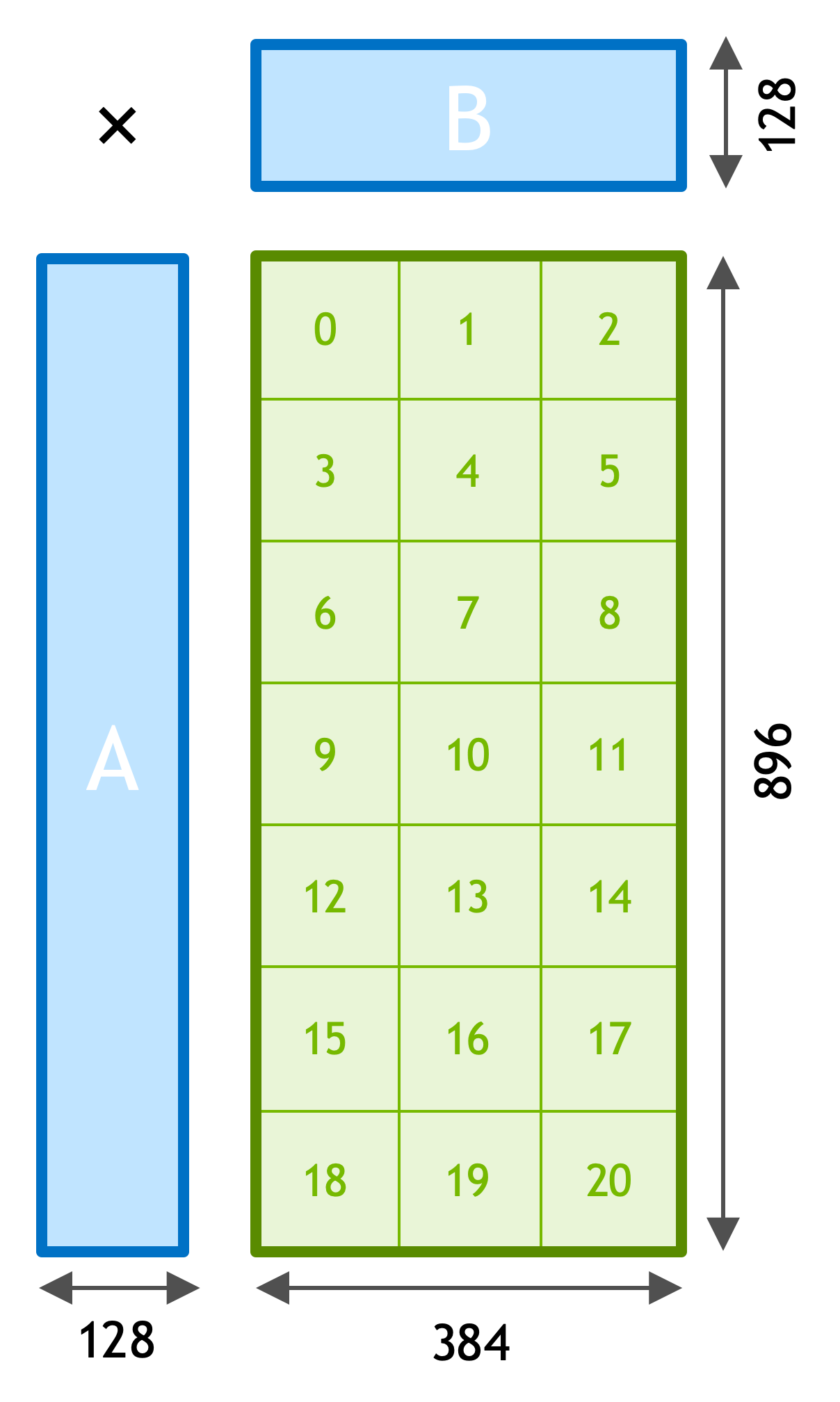}
    \end{subfigure}
    \hfill%
    \begin{subfigure}[t]{0.25\textwidth}
        \includegraphics[width=\linewidth]{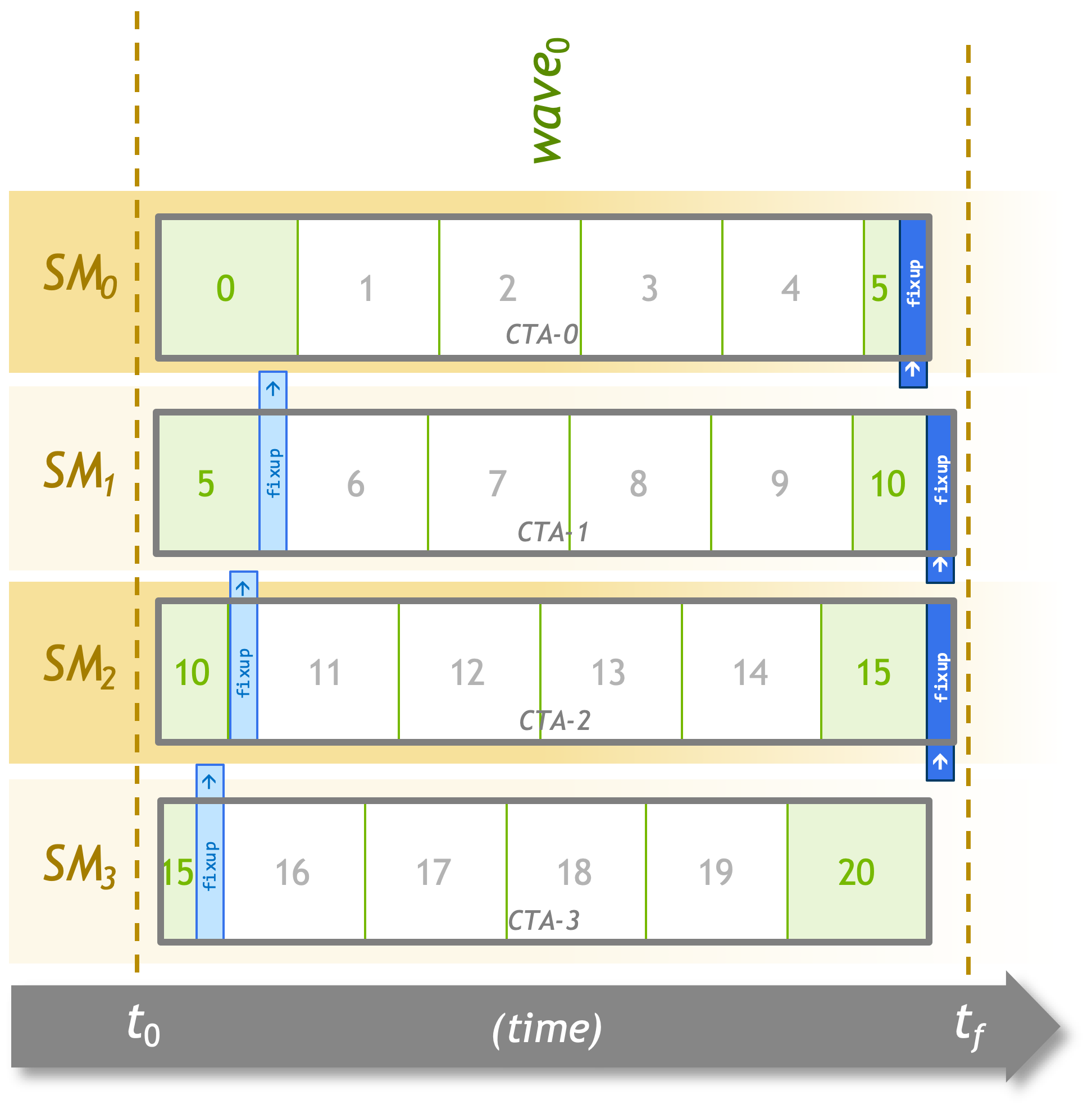}
        \caption{Basic Stream-K} \label{fig:basic_streamk}
    \end{subfigure}
    \hfill%
    \begin{subfigure}[t]{0.25\textwidth}
        \includegraphics[width=\linewidth]{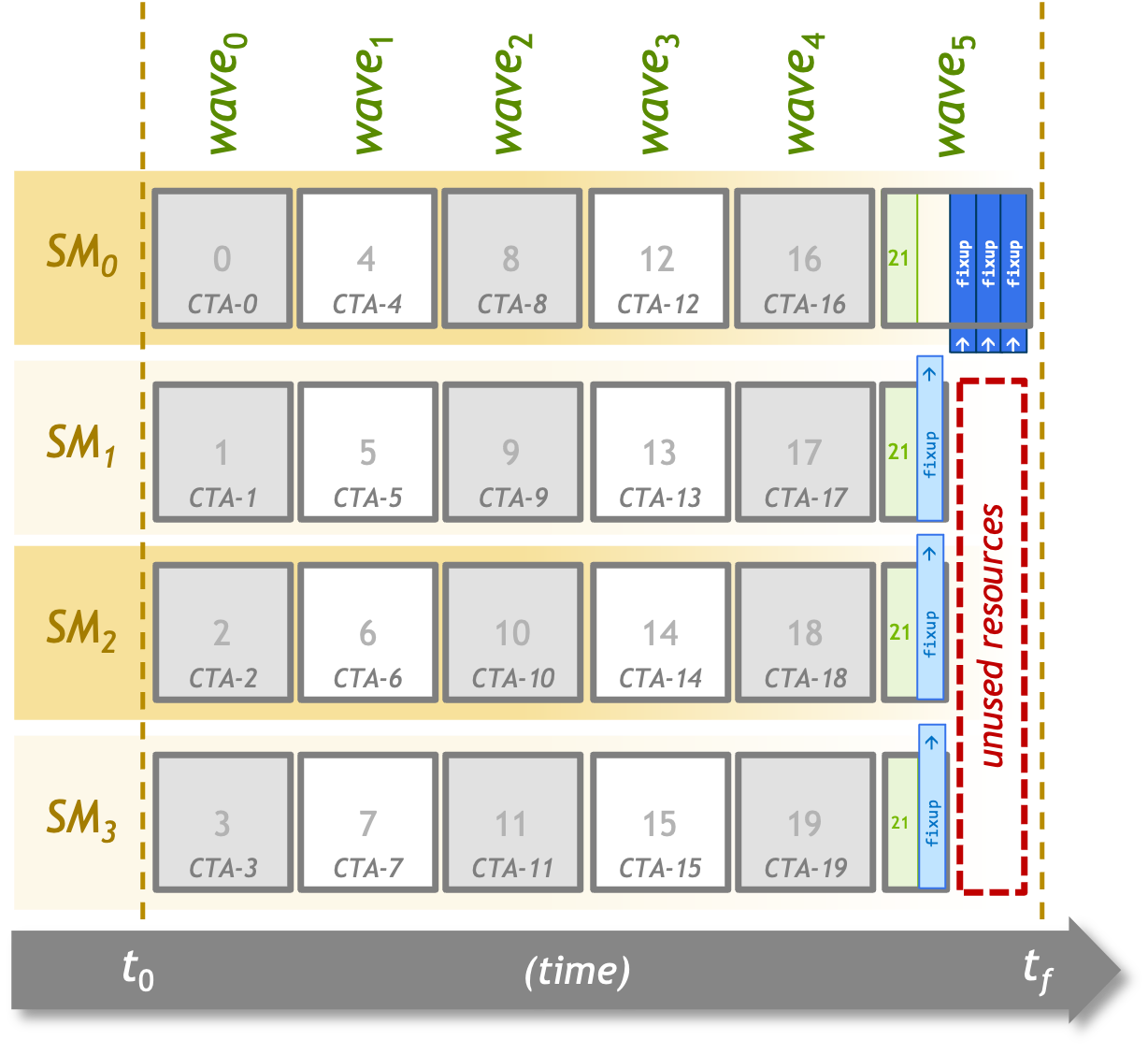}
        \caption{DP + one-tile SK} \label{fig:dp_one_tile_sk}
    \end{subfigure}
    \hfill%
    \begin{subfigure}[t]{0.25\textwidth}
        \includegraphics[width=\linewidth]{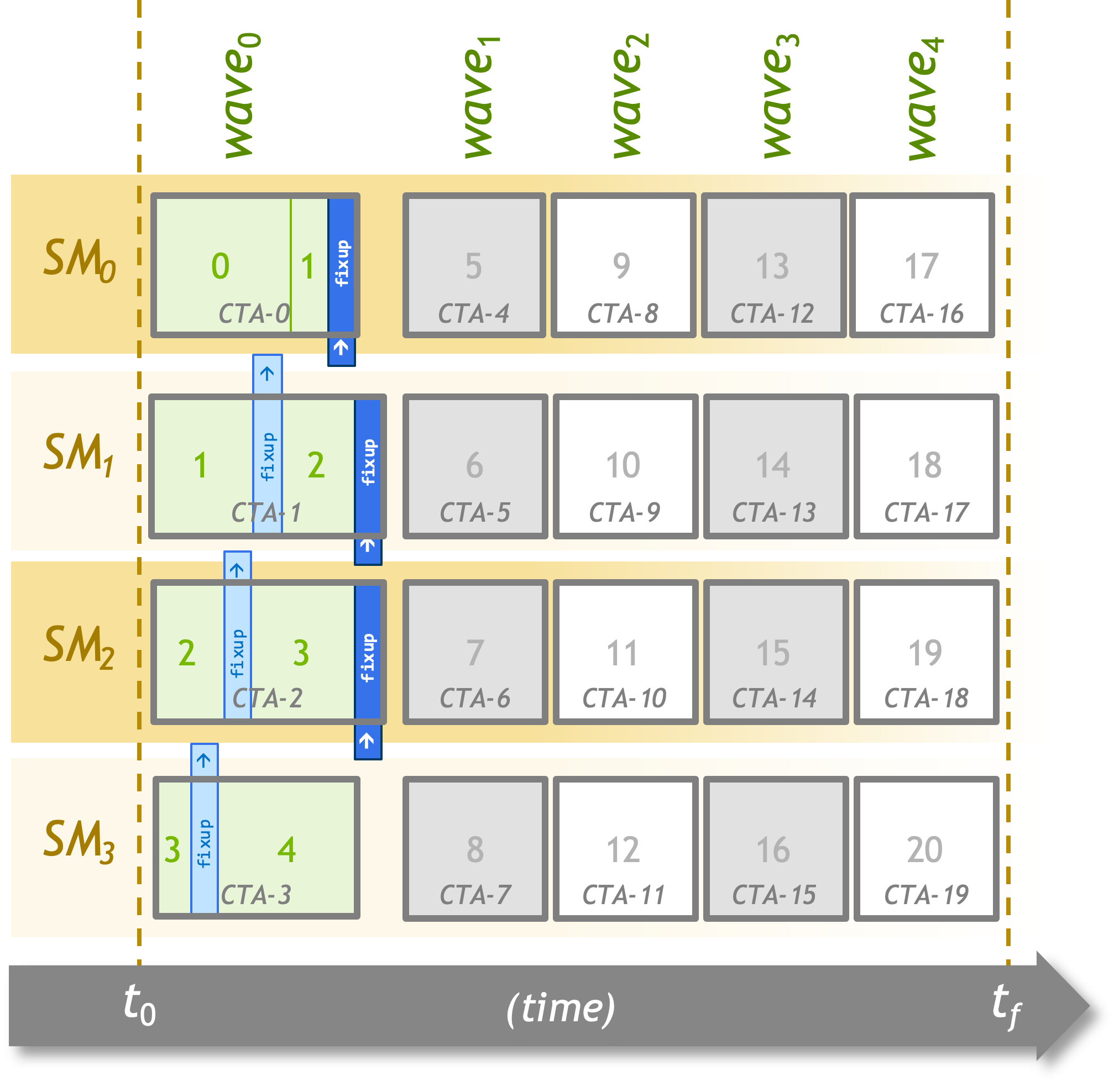}
        \caption{Two-tile SK + DP} \label{fig:two_tile_sk_dp}
    \end{subfigure}
    \caption{Basic \textit{Stream-K} vs.\ hybrid execution schedules for $896\times384\times128$ GEMM across a hypothetical four-SM GPU\@.}
\end{figure*}

The work decomposition we introduced in the last section can be
instantiated in a number of different ways to suit the needs of
different hardware architectures and software library designs.  Our
implementation targets NVIDIA GPUs and is designed to be integrated into
existing libraries like cuBLAS and CUTLASS.  In this section, we
describe how we configure the kernels we launch and introduce a
hybridization scheme that helps ensure users achieve maximum GEMM
performance across the widest possible range of problem shapes.

We also emphasize that these are truly internal implementation details.
They are completely transparent to the user of a BLAS-like library and
do not alter the library's interface.  The only observable impact is the
improved performance characteristics that we analyze in
Section~\ref{sec:evaluation}.

\subsection{Kernel Configuration}
\label{sec:blocking-factor-and-grid-configuration}

The tile size chosen for blocking the GEMM computation is, of course, a
critical parameter controlling the performance of the GEMM kernel.  For
modern NVIDIA GPUs, appropriate tile sizes are determined by the shape
of matrices supported by the GPU's Tensor Cores.  Based on extensive
empirical experience, we selected the smallest CTA-wide tile size
capable of achieving 99\% of the GPU's peak TFLOP/s for very large GEMM
volumes for each supported precision.  For the NVIDIA~A100 GPU used in
our experiments, these sizes are 64$\times$64$\times$16 for FP64
problems and 128$\times$128$\times$32 for FP16$\rightarrow$32 problems.

Achieving maximal GEMM performance from \emph{Stream-K} parallelization
also requires some degree of dynamic problem-specific configuration.
Before launching a kernel we choose a grid size likely to yield the best
performance on the specific problem shape at hand.  This is in contrast
to ensemble-based approaches which accommodate diverse problem shapes
through the static generation of many kernel variants based on workload
decomposition and blocking factor.

Our grid size selection heuristic is based on a simple analytical model
that minimizes the cost of reading, writing, and accumulating partial
sums while equally distributing the MAC-loop iterations per CTA.
Details of this analytical model are provided in the supplementary
material (Appendix~\ref{sec:analytical-modeling}).  Parameters to the
model are trivially chosen with empirical measurements and need only be
done once per target architecture.  The resulting parameters can then be
compiled statically into the library.  Again, this is in contrast to
ensemble-based approaches that rely on potentially complex
heuristics and machine learning models for kernel selection at run time.

\subsection{Data-parallel Hybridization}
\label{sec:data-parallel-hybridization}

The basic \emph{Stream-K} decomposition can, in certain cases, exhibit
tile-processing skew that leads to potentially adverse effects on cache
performance.
When
the number of output tiles $t$ is not an even multiple of the
grid size $g$, the starting $k$-offset for the first MAC-loop
iteration in each CTA will be different. Depending on the sizes and
shapes of the input matrices and blocking factors, this skew may
preclude these fragments from seeing reuse across CTAs in the GPU's
cache structure. In Figure~\ref{fig:basic_streamk}, for example, the initial $k$-axis
fragment offsets for each of the four CTAs will be $k=0$,
$k=32$, $k=64$, and $k=96$, respectively. Furthermore,
this 32-element skew between CTAs will persist for the duration of the
GEMM computation.

Tile-processing skew is a direct consequence of \emph{Stream-K}'s
workload balancing strategy.  However, we can take measures to limit its
duration by applying \emph{Stream-K}'s iteration balancing to a
smaller, tile-aligned region of the total iteration domain such that the
remaining tiles can be produced in full, temporally aligned waves.

The simplest hybrid scheme is the
``\emph{data-parallel} + one-tile \emph{Stream-K}''
schedule illustrated in Figure~\ref{fig:dp_one_tile_sk}.  It applies iteration balancing
only among the tiles otherwise remaining for a final, partially full
\emph{data-parallel} wave.
The total number of full waves is
$w = \lfloor t/p \rfloor$, where $t$ is the number of output
tiles and $p$ is the number of SM cores in the GPU\@. Consequently,
each \emph{Stream-K} CTA receives an even share of iterations that is
less than one tile's worth.
Unfortunately, this strategy has little ability to hide the
synchronization latency for the exchange of partial sums when three or
more CTAs cover the same tile. In these scenarios, the accumulating CTA
may be forced to wait for the contributions of other CTAs to become
visible, as all but the last will be completing their final iterations
at roughly the same time. Furthermore, the basic version of our scheme 
for aggregating partials is serialized within a single CTA, and thus will 
likely cause SM workload imbalance when the number of contributing CTAs 
per tile is large.

We address these problems with
our ``two-tile \emph{Stream-K} + \emph{data-parallel}'' hybrid schedule,
illustrated in Figure~\ref{fig:two_tile_sk_dp}.
It performs
one fewer full data-parallel wave in exchange for each \emph{Stream-K}
CTA receiving more than one tile's worth of iterations (but fewer than
two). This provides much better latency hiding when $w\geq2$,
and each accumulating CTA will only need to receive partials from
one other contributing CTA\@. Otherwise, it behaves identically to the
``\emph{DP + one tile SK}'' schedule.
This hybrid approach results in both improved memory access patterns and latency hiding.
It also shows the versatility of the generic \emph{Stream-K} looping structure to implement different scheduling policies within the same kernel instance.

\section{Performance Evaluation} \label{sec:evaluation}

We have implemented our \emph{Stream-K} decomposition using NVIDIA's
CUTLASS library of CUDA C++ template abstractions for authoring
GEMM-like computations. CUTLASS provides the optimized equivalent
of the CTA-wide \lstinline{MacLoop()} subroutine in Algorithm~\ref{alg:macloop},
which performs blocking, tiling, and software-pipelined
data movement that is analogous to the closed-source cuBLAS and cuDNN
implementations.
Our evaluation encompasses both (1) double-precision FP64 GEMM, and
(2) mixed-precision FP16$\rightarrow$32 GEMM. For the latter, the input
matrices \textbf{A} and \textbf{B} comprise half-precision FP16 values,
yet the internal accumulation and output matrix \textbf{C} values are
single-precision FP32.

\paragraph{Hardware environment.} Our test GPU is the NVIDIA A100, which contains 108~SM cores.
For measurement
stability, we lock the power envelope at 400~W and SM clocks at 1005~MHz
($\sim$71\% of their dynamic peak).  This establishes FP64 tensor-core
peak throughput of 13.9~TFLOP/s, and mixed FP16$\rightarrow$32
tensor-core peak throughput of 222.3 TFLOP/s.

\begin{figure}
    \centering
    \includegraphics[width=\columnwidth]{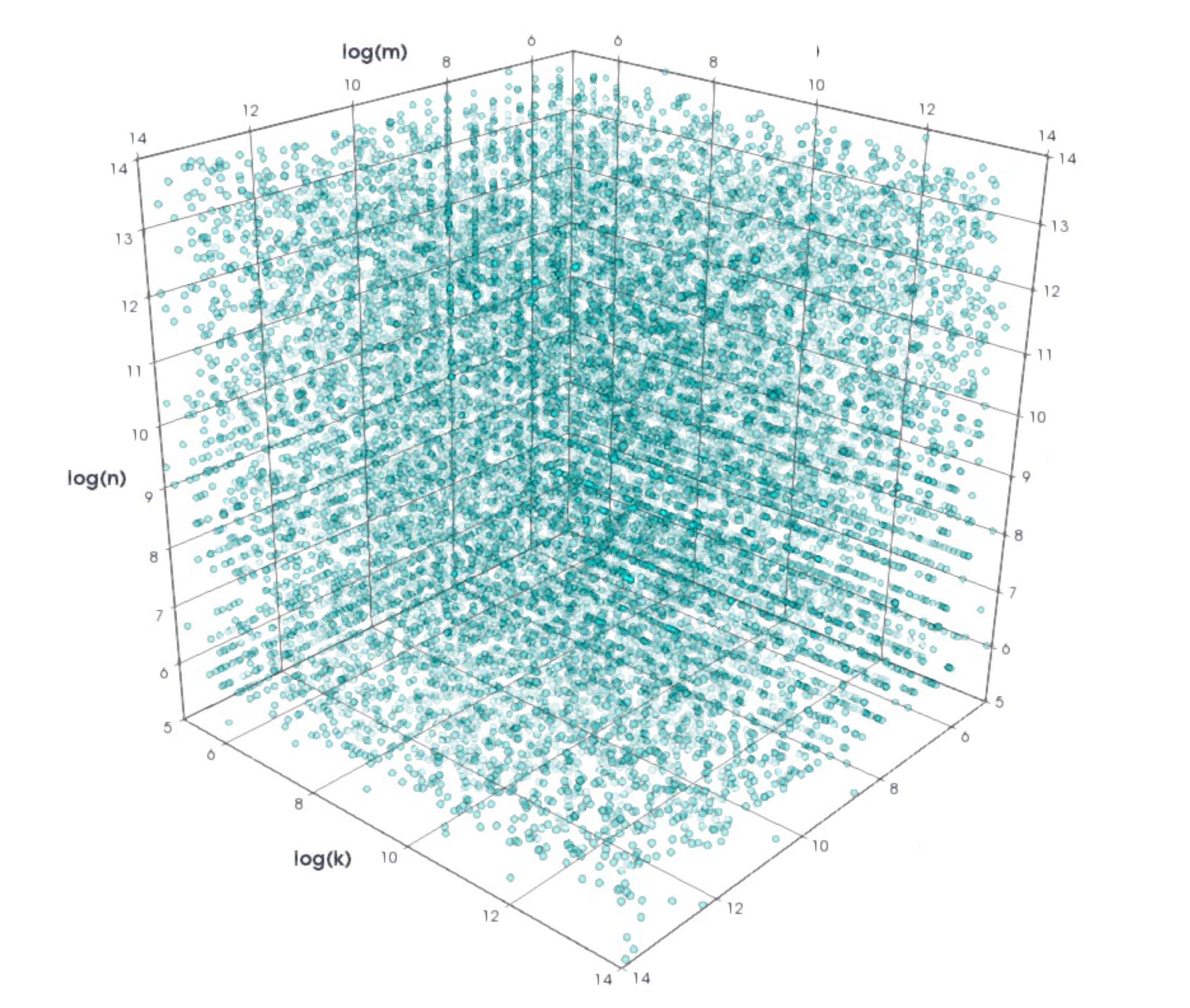}
    \caption{The test domain of 32,824 GEMM problem shapes and sizes used for performance evaluation.\\
        $\{m\} = \{128 \ldots8192\}$,~
        $\{n\} = \{128 \ldots8192\}$,~
        $\{k\} = \{128 \ldots8192\}$} \label{fig:test_corpus}
\end{figure}

\paragraph{Dataset.} Our test corpus intends to approximate the enormous breadth
and scope of device-wide GEMM problems that GPU math kernel libraries are
designed to accommodate. As shown in Figure~\ref{fig:test_corpus}, we evaluate
32,824 different problem sizes and shapes, log-sampled at random within a
domain of $m$, $n$, and $k$ matrix dimensions whose volume spans six orders
of magnitude.

\paragraph{Methodology.} For both GEMM precisions,
we build a single \emph{Stream-K} kernel
that has been specialized per the guidelines in the Section~\ref{sec:practical-usage}.
Furthermore, these kernels implement our ``two-tile \emph{Stream-K} +
\emph{data-parallel}'' hybrid decomposition.
Our evaluation compares each \emph{Stream-K} kernel with:
\begin{enumerate}
    \item the default \emph{data-parallel} CUTLASS kernel of the same blocking factor;
    \item the cuBLAS ensemble for that precision (CUDA 11.6); and
    \item an idealized oracle that will always select the highest
          performing \emph{data-parallel} CUTLASS blocking factor to execute
          for a given GEMM instance.
\end{enumerate}
For FP64 problems, this oracle
selects among the ensemble of \{(32$\times$32$\times$16), (32$\times$64$\times$16),
(64$\times$64$\times$16), (64$\times$128$\times$16), (128$\times$128$\times$16)\}
blocking factor specializations.  For FP16$\rightarrow$32, it selects
among the ensemble of \{(64$\times$64$\times$64), (64$\times$128$\times$32),
(128$\times$128$\times$32), (128$\times$256$\times$32)\} blocking factor
specializations. These specific specializations
are an open-sourced strict subsets alternative of the corresponding
cuBLAS GEMM kernel ensembles.

\begin{figure*}
    \centering
    \begin{subfigure}[t]{\columnwidth}
        \includegraphics[width=\columnwidth]{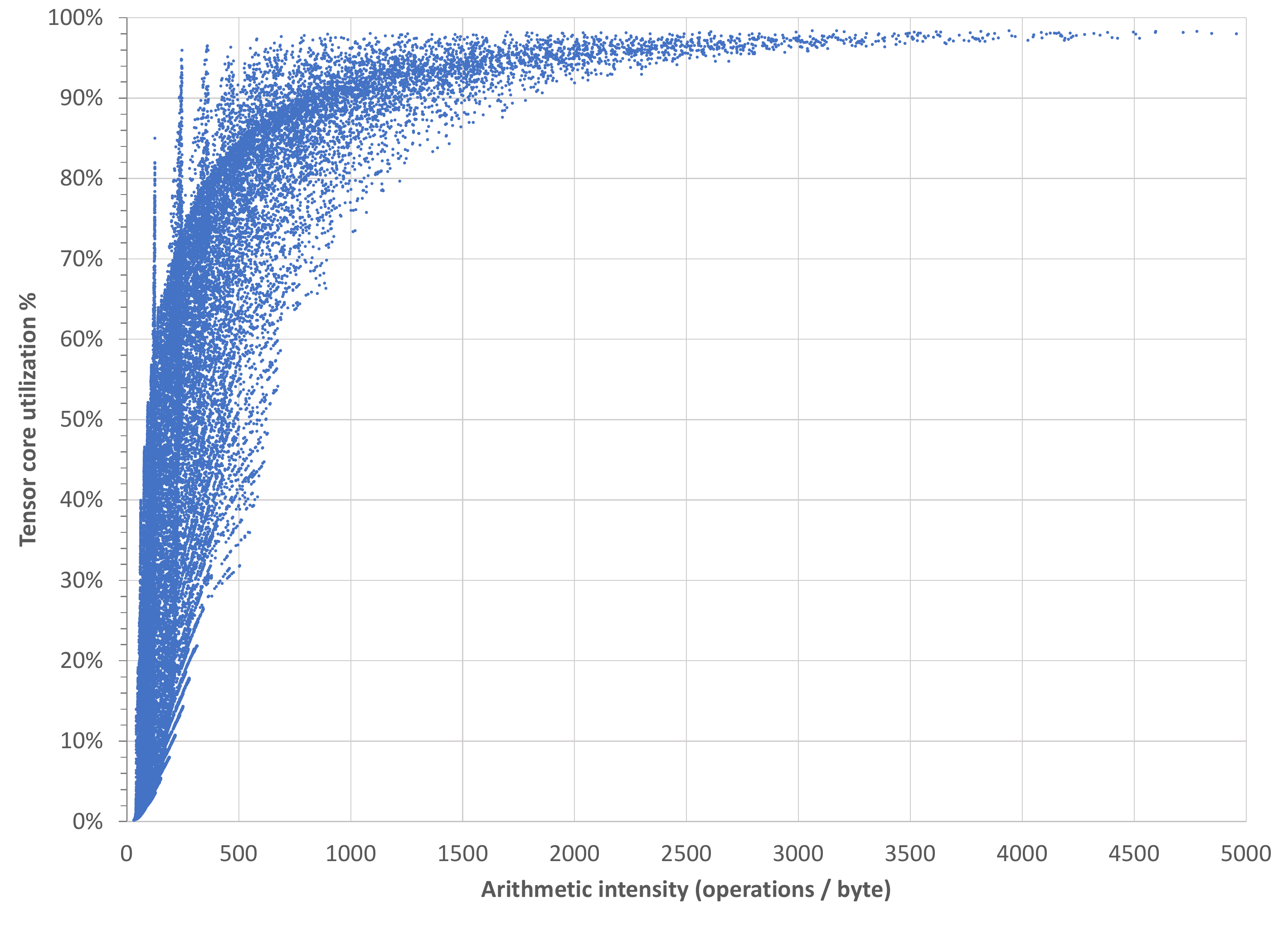}
        \caption{CUTLASS FP16$\rightarrow$32 \emph{data-parallel} ``roofline''\\
            performance ($\text{blocking factors} = $ 128$\times$128$\times$32).} \label{fig:hgemm_roofline_cutlass}
    \end{subfigure}
    \hfill%
    \begin{subfigure}[t]{\columnwidth}
        \includegraphics[width=\columnwidth]{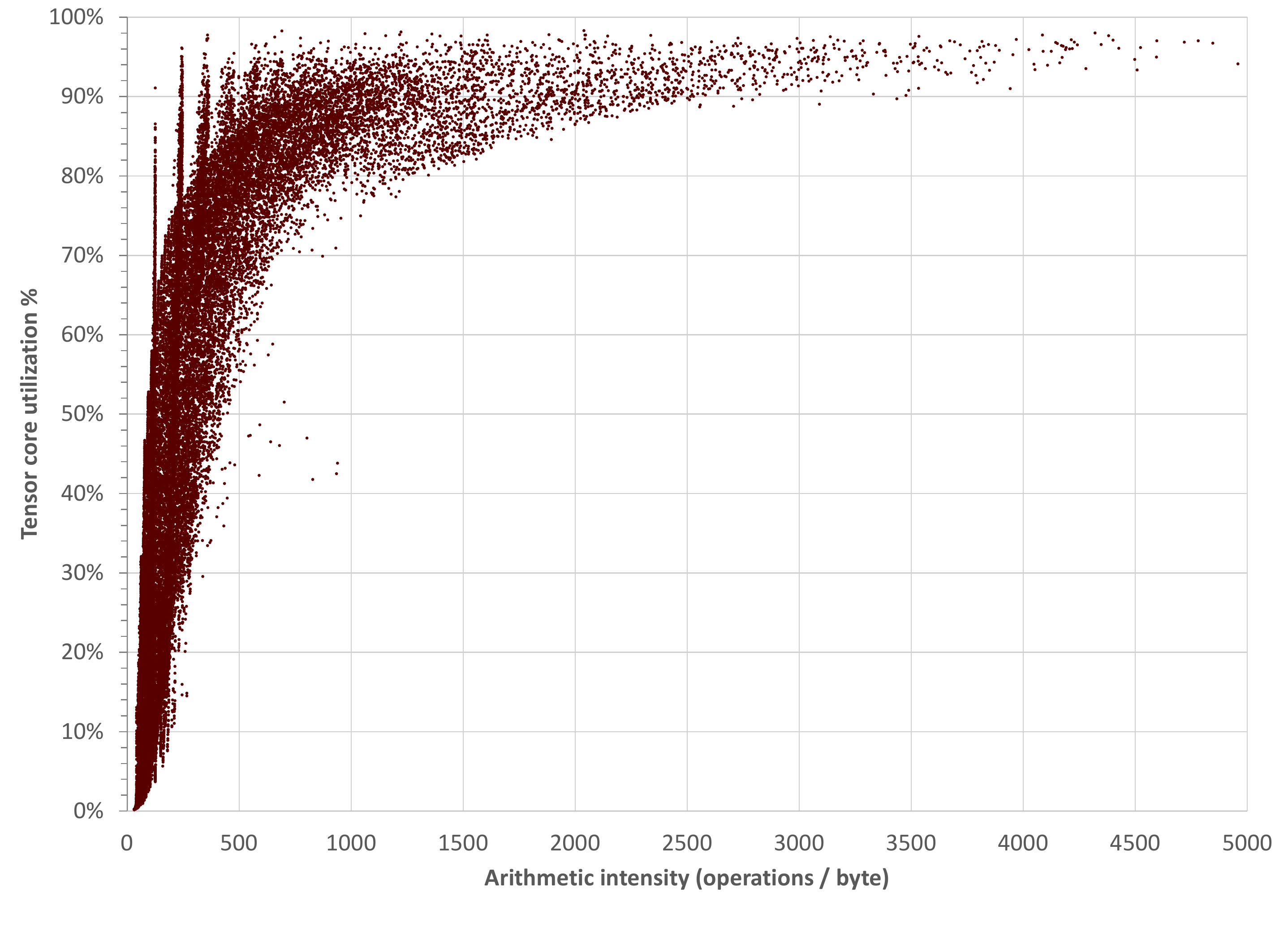}
        \caption{cuBLAS
            (ensemble)} \label{fig:hgemm_roofline_cublas}
    \end{subfigure}
    \vfill%
    \begin{subfigure}[t]{\columnwidth}
        \includegraphics[width=\columnwidth]{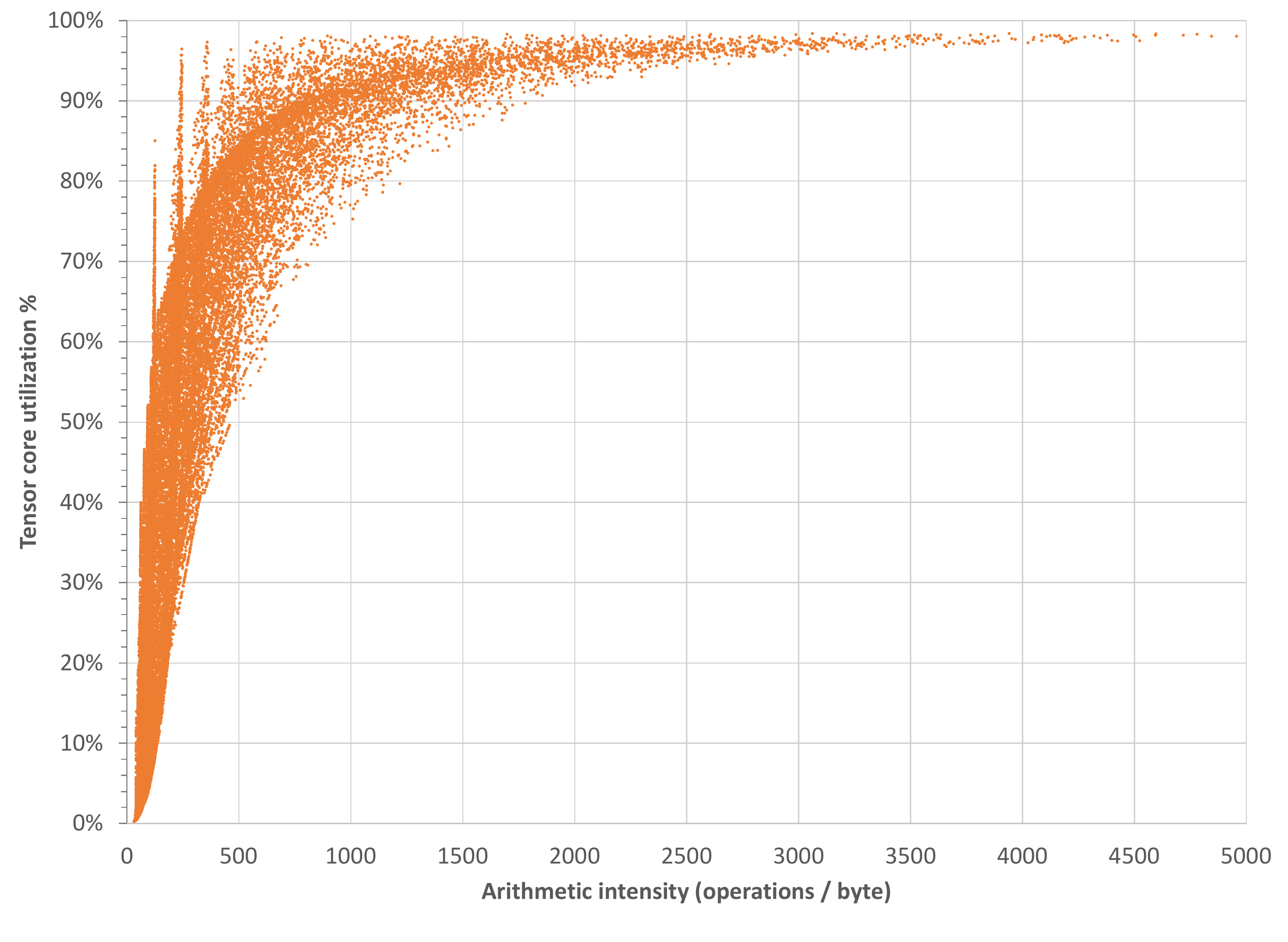}
        \caption{Idealized CUTLASS oracle
            (ensemble)} \label{fig:hgemm_roofline_oracle}
    \end{subfigure}
    \hfill%
    \begin{subfigure}[t]{\columnwidth}
        \includegraphics[width=\columnwidth]{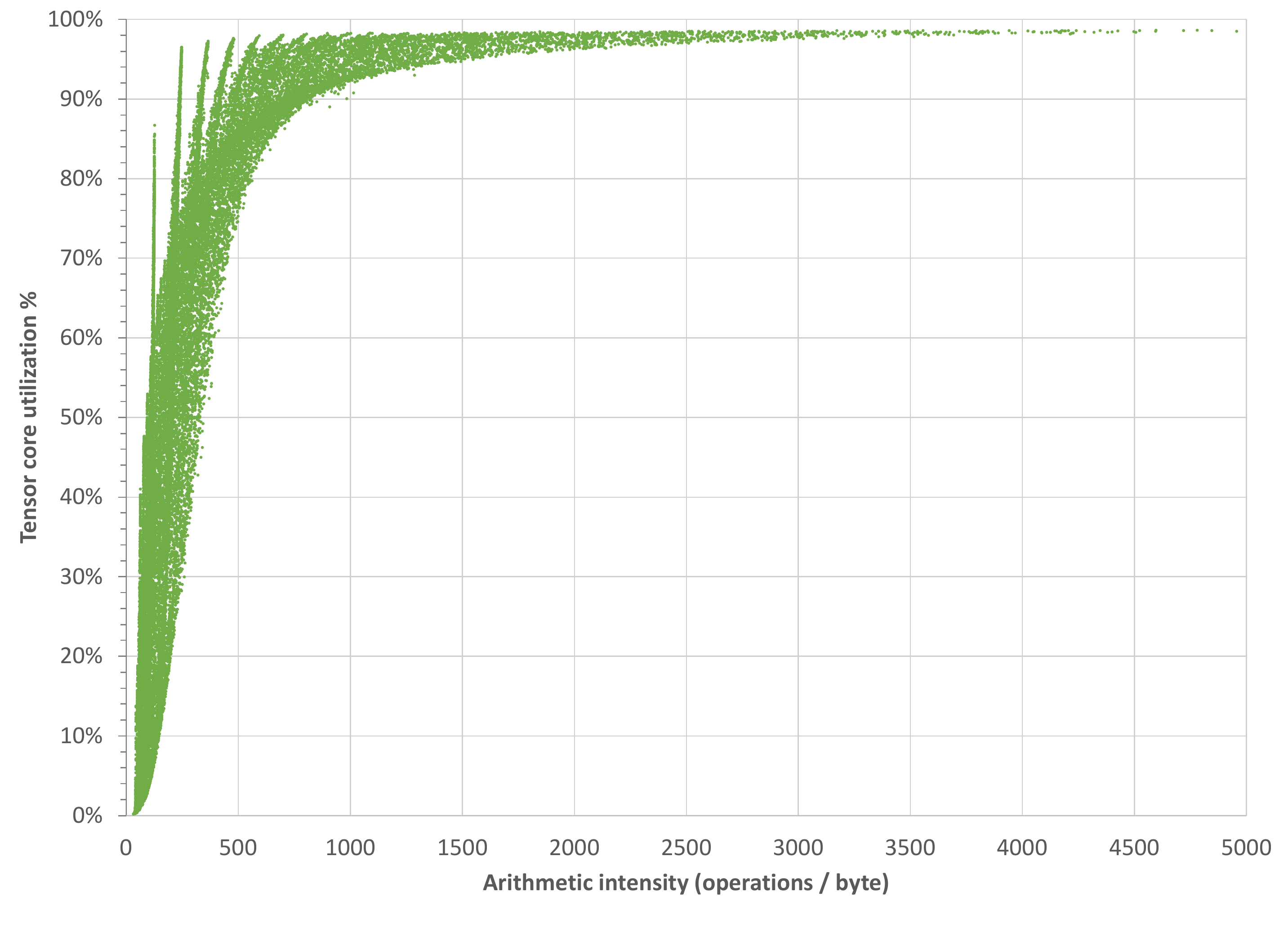}
        \caption{\emph{Stream-K}
            ($\text{blocking factors} = $ 128$\times$128$\times$32)} \label{fig:hgemm_roofline_streamk}
    \end{subfigure}
    \caption{FP16$\rightarrow$FP32 GEMM ``roofline'' performance utilization landscapes on NVIDIA A100 across 32K GEMM problem shapes and sizes.} \label{fig:hgemm_comparison}
\end{figure*}

\begin{figure*}
    \centering
    \begin{subfigure}[t]{\columnwidth}
        \includegraphics[width=\columnwidth]{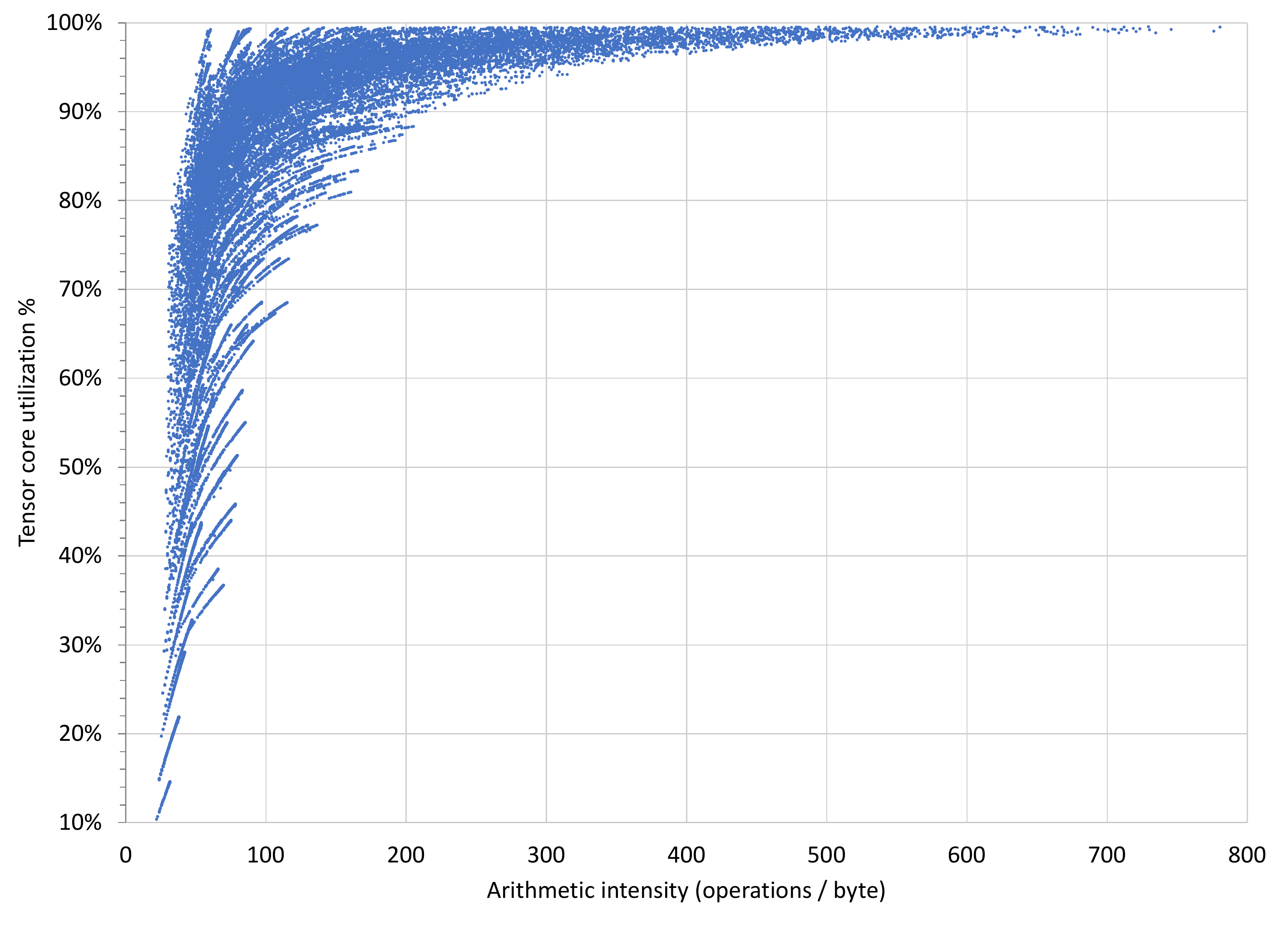}
        \caption{CUTLASS \emph{data-parallel} \\
            ($\text{blocking factors} = $ 64$\times$64$\times$16)} \label{fig:dgemm_roofline_cutlass}
    \end{subfigure}
    \hfill%
    \begin{subfigure}[t]{\columnwidth}
        \includegraphics[width=\columnwidth]{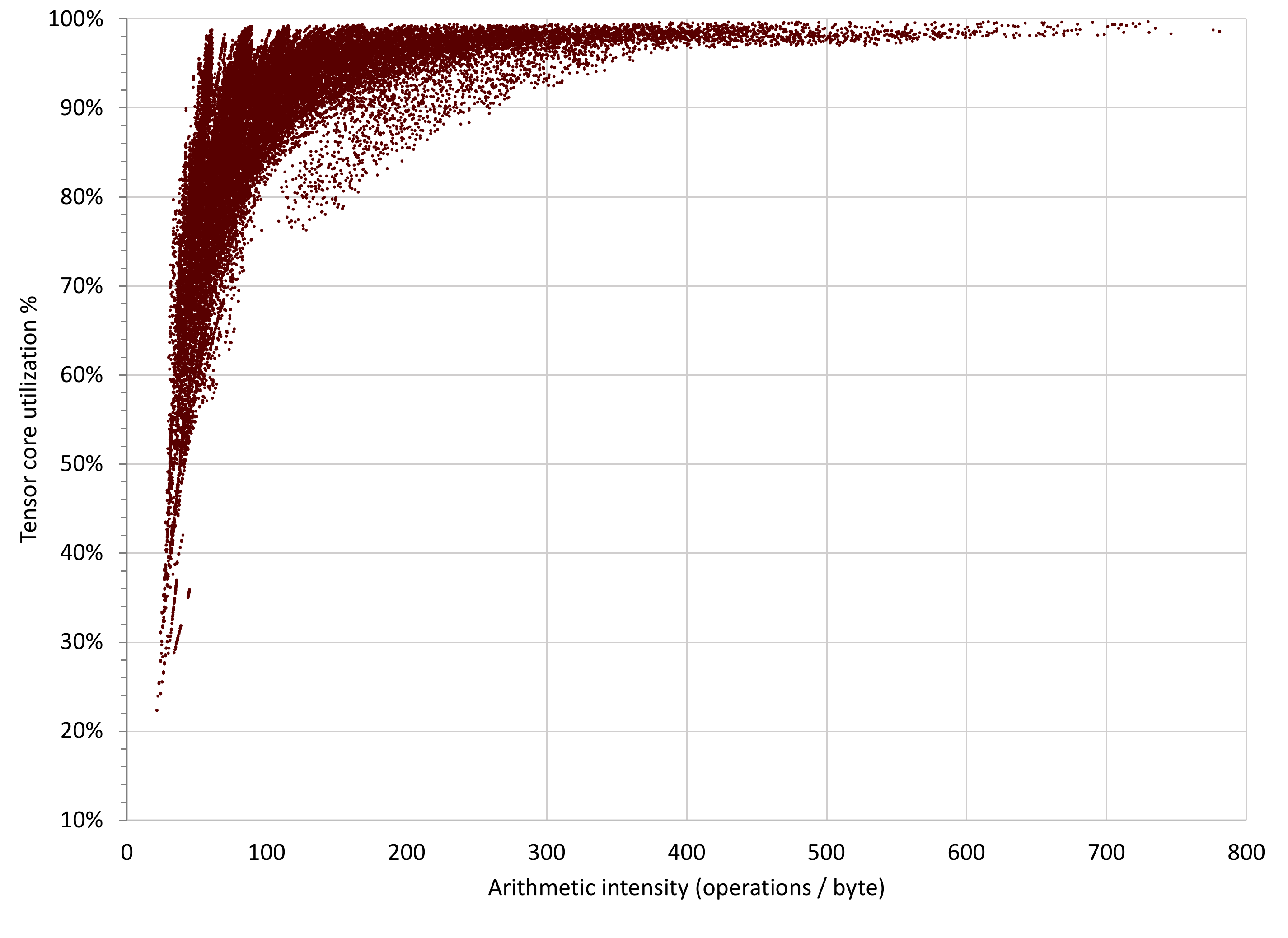}
        \caption{cuBLAS
            (ensemble)} \label{fig:dgemm_roofline_cublas}
    \end{subfigure}
    \vfill%
    \begin{subfigure}[t]{\columnwidth}
        \includegraphics[width=\columnwidth]{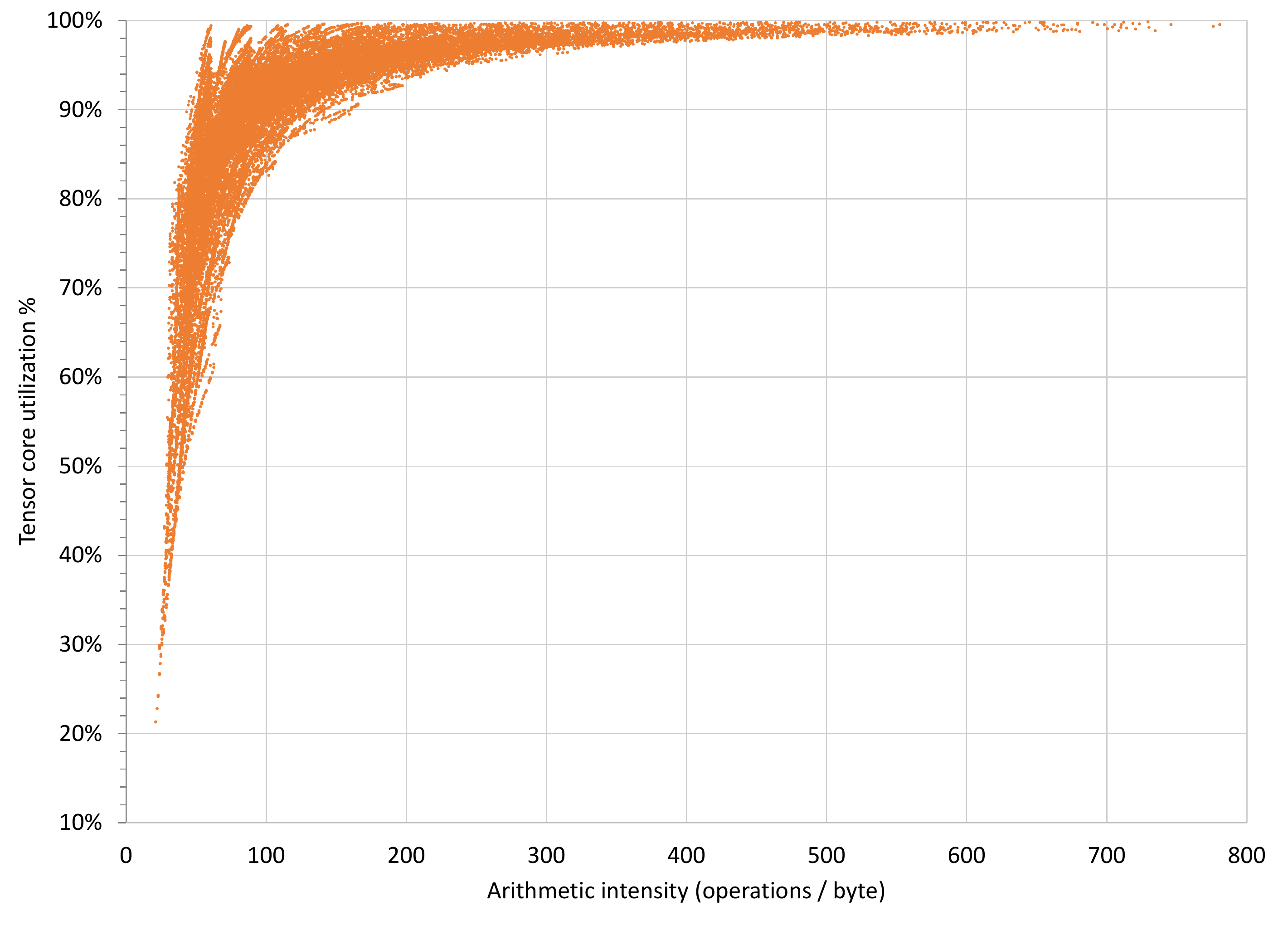}
        \caption{Idealized CUTLASS oracle
            (ensemble)} \label{fig:dgemm_roofline_oracle}
    \end{subfigure}
    \hfill%
    \begin{subfigure}[t]{\columnwidth}
        \includegraphics[width=\columnwidth]{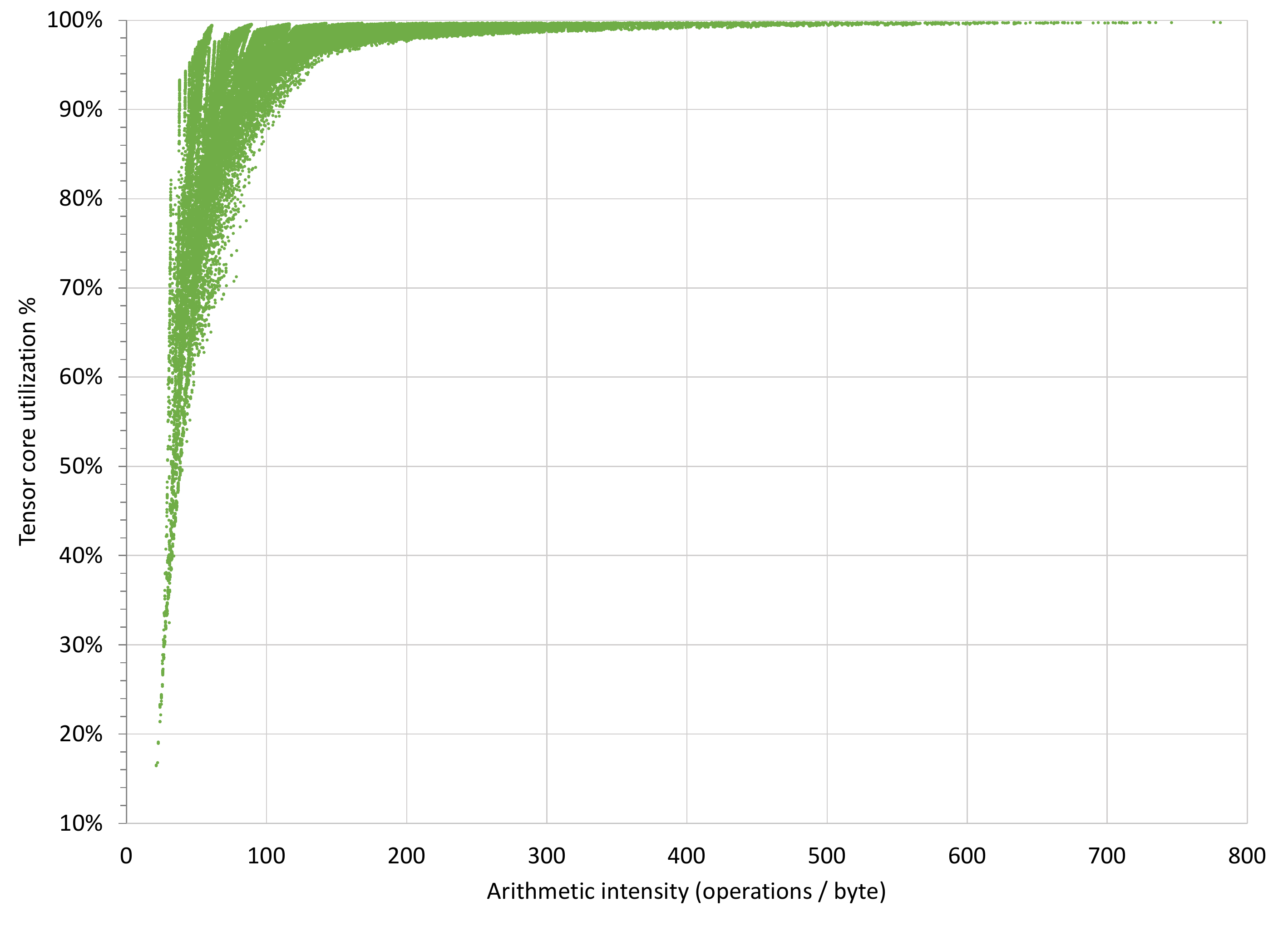}
        \caption{\emph{Stream-K}
            ($\text{blocking factors} = $ 64$\times$64$\times$16)} \label{fig:dgemm_roofline_streamk}
    \end{subfigure}
    \caption{FP64 GEMM ``roofline'' performance utilization landscapes on NVIDIA A100 across 32K problem shapes and sizes.} \label{fig:dgemm_comparison}
\end{figure*}

\begin{table}
    \footnotesize
    \begin{tabularx}{\columnwidth}{rXXXX}
        \toprule
                         & \thead{vs.\                                               \\ CUTLASS\\ {\tiny $64\times64\times16$}}
                         & \thead{vs.\                                               \\ cuBLAS}
                         & \thead{vs.\                                               \\ cuBLAS\\ {\tiny $>150$ ops/B}}
                         & \thead{vs.\                                               \\ CUTLASS \\ oracle} \\
        \midrule
        \textbf{Average} & $1.23\times$ & $1.06\times$ & $1.03\times$ & $1.05\times$ \\
        \textbf{StdDev}  & $0.45$       & $0.10$       & $0.03$       & $0.09$       \\
        \textbf{Min}     & $0.77\times$ & $0.68\times$ & $0.99\times$ & $0.70\times$ \\
        \textbf{Max}     & $5.63\times$ & $2.55\times$ & $1.24\times$ & $1.64\times$ \\
        \bottomrule
    \end{tabularx}
    \caption{\emph{Stream-K} FP64 Relative Performance}
    \label{tab:streamk_speedup_fp64}

    \begin{tabularx}{\columnwidth}{rXXXX}
        \toprule
                         & \thead{vs.\                                               \\ CUTLASS\\ {\tiny $128\times128\times32$}}
                         & \thead{vs.\                                               \\ cuBLAS}
                         & \thead{vs.\                                               \\ cuBLAS\\ {\tiny $>150$ ops/B}}
                         & \thead{vs.\                                               \\ CUTLASS \\ oracle} \\
        \midrule
        \textbf{Average} & $1.63\times$ & $1.13\times$ & $1.15\times$ & $1.12\times$ \\
        \textbf{StdDev}  & $1.46$       & $0.45$       & $0.12$       & $0.37$       \\
        \textbf{Min}     & $0.80\times$ & $0.64\times$ & $0.98\times$ & $0.61\times$ \\
        \textbf{Max}     & $14.7\times$ & $6.74\times$ & $1.85\times$ & $4.63\times$ \\
        \bottomrule
    \end{tabularx}
    \caption{\emph{Stream-K} FP16$\rightarrow$32 Relative Performance}
    \label{tab:streamk_speedup_fp16}
\end{table}

The ``roofline'' plots of Figure~\ref{fig:dgemm_roofline_cutlass} and
Figure~\ref{fig:hgemm_roofline_cutlass} highlight the spread of
performance produced by the singleton \emph{data-parallel} CUTLASS kernels.
They plot the percentage of FP64 and FP16$\rightarrow$32 processor utilization
as a function of computational intensity. Ideally, a GEMM implementation's
performance response would manifest as a narrow band that adheres tightly to
the machine's bandwidth- and compute-bound performance ceilings. Here, the
\emph{data-parallel} kernels exhibit a fairly large dynamic range for any given
regime of arithmetic intensity.  In contrast, the performance responses from
the equivalent \emph{Stream-K} kernels in Figure~\ref{fig:dgemm_roofline_streamk}
and Figure~\ref{fig:hgemm_roofline_streamk} are much tighter. These
observations are corroborated by Table~\ref{tab:streamk_speedup_fp64}
and Table~\ref{tab:streamk_speedup_fp16}, which show the \emph{Stream-K}
kernels outperforming their \emph{data-parallel} FP64 and FP16$\rightarrow$32
equivalents by an average of 1.23$\times$ and 1.63$\times$, respectively.
For extreme strong-scaling scenarios where $m \times n$ is small and $k$ is
large, our \emph{Stream-K} kernels demonstrate up to 5.63$\times$ and
14.7~$\times$ speedup, respectively.

The second columns of Table~\ref{tab:streamk_speedup_fp64} and
Table~\ref{tab:streamk_speedup_fp16} compare our \emph{Stream-K} performance
with that of cuBLAS\@. On average, our FP64 and FP16$\rightarrow$32
\emph{Stream-K} GEMM kernels respectively deliver 6\% and 13\% greater
throughput than their corresponding cuBLAS ensembles, with peak improvement of 2.55$\times$ and 6.74$\times$. This is a significant improvement over the breadth of 32K GEMM problem shapes and sizes with 20$\times$ \emph{less} executable code (a single kernel for each precision) than NVIDIA's vendor GEMM library, cuBLAS\@.


Furthermore, the contrast between the FP64 and FP16$\rightarrow$32
cuBLAS performance responses (Figure~\ref{fig:dgemm_roofline_cublas} and
Figure~\ref{fig:hgemm_roofline_cublas}) versus those of our hypothetical CUTLASS
oracle ensembles (Figure~\ref{fig:dgemm_roofline_oracle} and
Figure~\ref{fig:hgemm_roofline_oracle}) reveal the difficulties of designing
kernel selection heuristics that deliver consistently good performance.
Despite having access to the same blocking factor specializations, cuBLAS
exhibits substantially wider dynamic ranges than the idealized
\emph{data-parallel} CUTLASS oracle. The performance spreads of our
\emph{Stream-K} kernels are narrower still, achieving up to 4.6$\times$
the idealized oracle performance and underscoring their ability to achieve
utilization levels that are simply not possible from tile-centric
work decompositions.

\begin{figure*}
    \begin{subfigure}[t]{\columnwidth}
        \centering
        \includegraphics[width=\columnwidth]{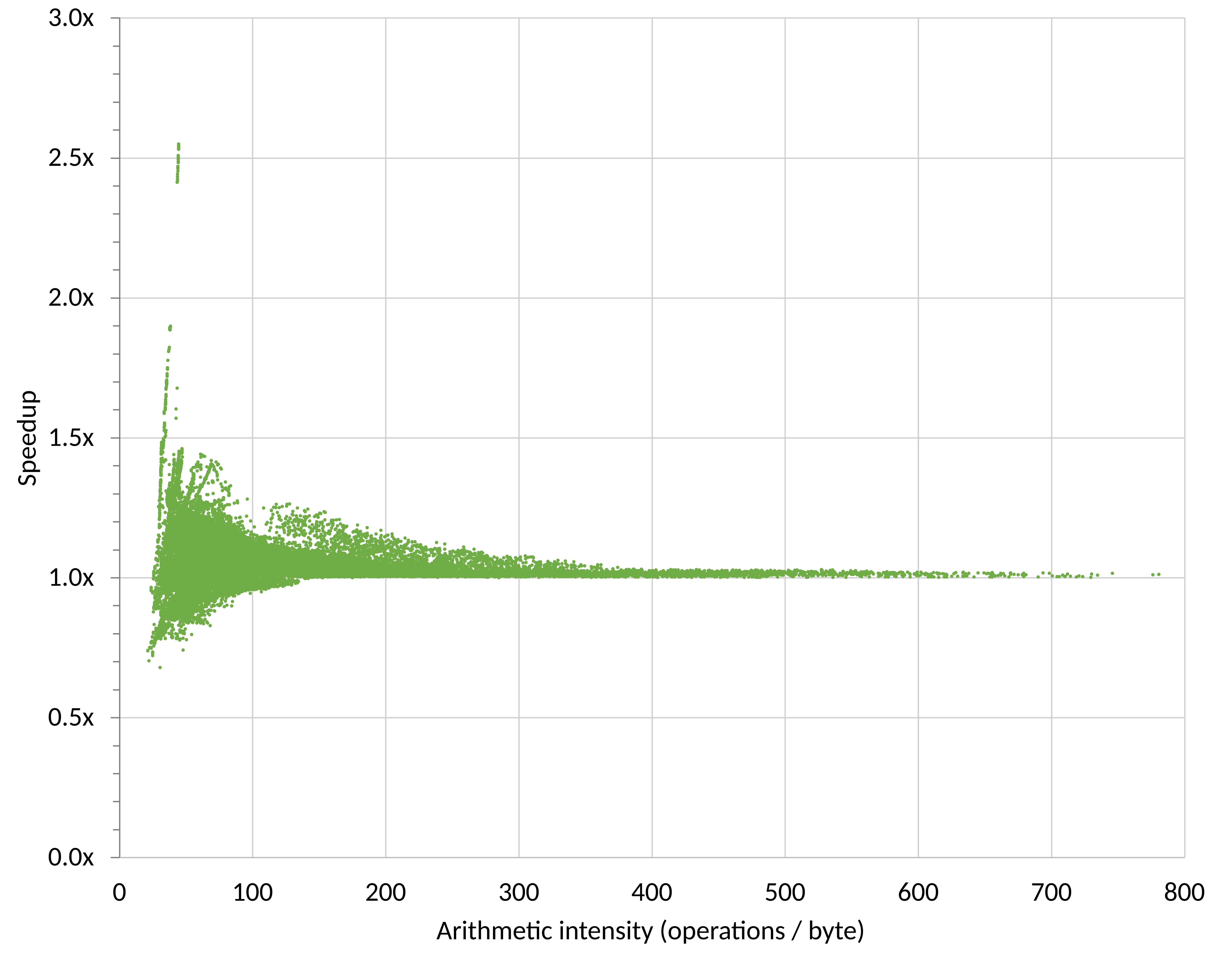}
        \caption{FP64 \emph{Stream-K} speedup vs.\ cuBLAS.} \label{fig:dgemm_speedup}
    \end{subfigure}
    \hfill%
    \begin{subfigure}[t]{\columnwidth}
        \centering
        \includegraphics[width=\columnwidth]{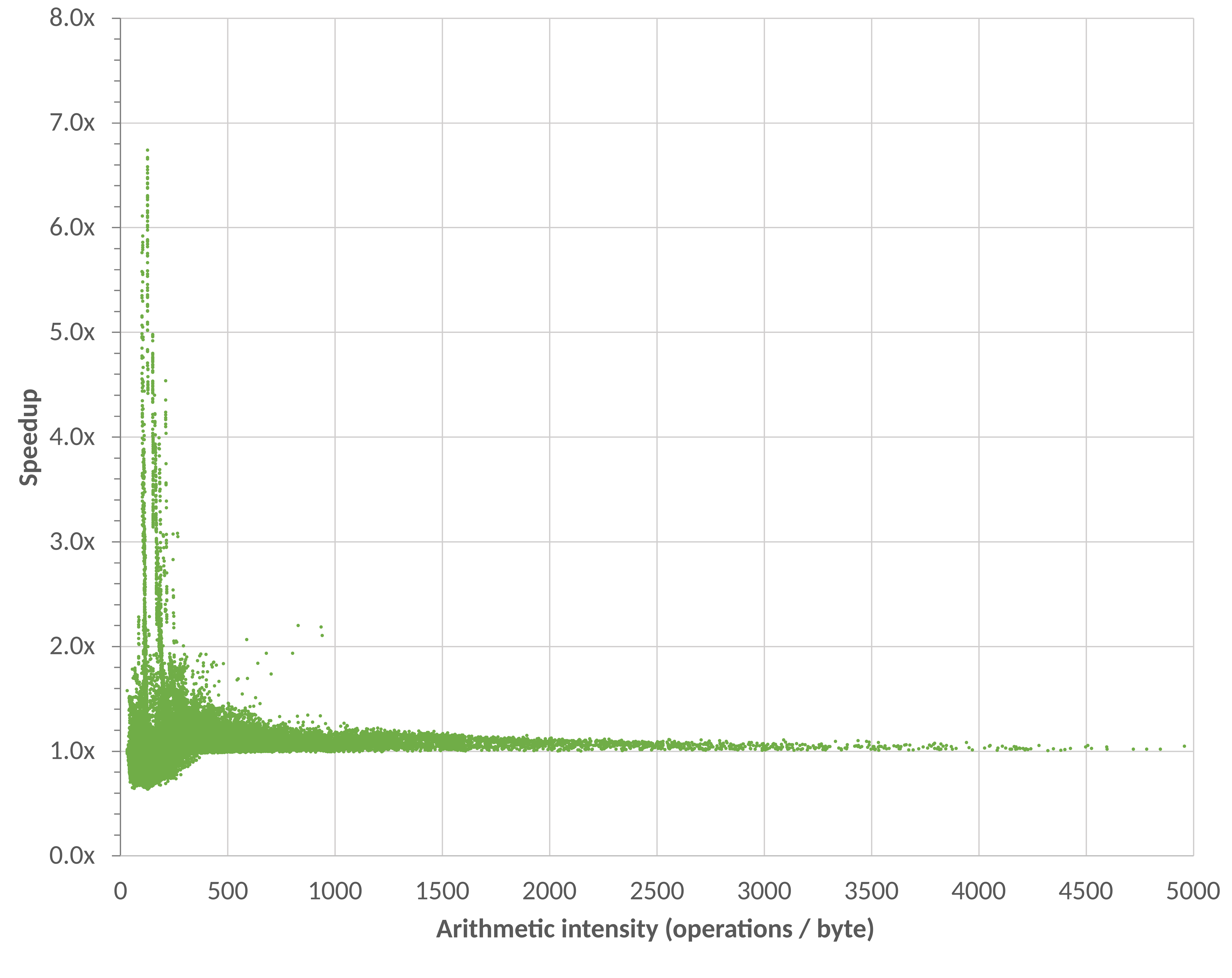}
        \caption{FP16$\rightarrow$32 \emph{Stream-K} speedup vs.\ cuBLAS.} \label{fig:hgemm_speedup}
    \end{subfigure}
    \caption{\emph{Stream-K} speedup vs.\ cuBLAS.} \label{fig:cublas_speedup}
\end{figure*}

Finally, we observe regimes of small, bandwidth-bound problem shapes
where our largish blocking factors do not compete well against cuBLAS.
However, if we restrict our scope to the domain of compute-bound problems
(i.e., FP64 problems having compute intensity $>$~150~ops/byte and
FP16 $\rightarrow$ 32 problems $>$~400~ops/byte), Figure~\ref{fig:dgemm_speedup}
and Figure~\ref{fig:hgemm_speedup} demonstrate that our singleton
\emph{Stream-K} kernels achieve unilaterally higher performance than the
cuBLAS ensembles.
The ``noisy'' relative performance in the regimes below these thresholds is not
surprising, as \emph{Stream-K} is attempting to make memory-bound
computations run faster by adding more memory workload. This suggests a
few avenues for future work, namely separate cost-modeling for the
memory-bound regime and/or the bundling of a second \emph{Stream-K} kernel
having smaller tile size into a two-kernel ensemble.

\section{Conclusion}

We presented \emph{Stream-K}, a novel parallel workload decomposition
technique for scheduling general matrix multiplication (GEMM) and
similar computations on wide architectures such as GPUs. Unlike other
tile-splitting techniques, the MAC-loop iteration is our unit of
workload quantization across processor cores. This affords excellent
strong scaling and workload balancing because its cost is (1) a constant
with respect to the problem shape, and (2) substantially smaller than
that of an entire output tile.

Furthermore, \emph{Stream-K} produces an $O(p)$ number of splitting
seams that are bound by the number of processor cores. Consequently, the
overheads of strong scaling and workload balancing scale with processor
width rather than problem size. This is a welcome feature for many
applications that cannot afford to allocate large amounts of temporary storage
equivalent to the problem output.

Finally, we evaluated our \emph{Stream-K} approach across a broad spectrum 
of GEMM shapes and sizes.  We showed that a single blocking configuration of
\emph{Stream-K} can (1) achieve levels of absolute performance that match 
and/or exceed that of NVIDIA's cuBLAS library, even when the latter is 
operating at near-peak processor utilization, and (2) do so with much 
higher levels of performance consistency. Additionally, \emph{Stream-K} 
is an attractive option for library construction and maintenance, as 
it presents an opportunity to reduce distribution sizes by an order 
of magnitude and removes the need for complex handcoded heuristics or
machine learning models for kernel selection without compromising performance.
\emph{Stream-K} is open-sourced within CUTLASS~2.11 (\url{https://github.com/NVIDIA/cutlass}) and the performance shown within this paper can be reproduced when compiled using CUDA~11.8.

For future works, we identify cache-aware, tile-access patterns such as 
Morton Order, an avenue for optimization. We also believe that 
\emph{Stream-K} decomposition could provide a similar improved performance 
response for other GEMM-like workloads that struggle with the same 
quantization inefficiencies.


\begin{acks}
    This material is based upon work supported by \grantsponsor{}{Defense Advanced Research Projects Agency (DARPA)}{} under Contract No.~\grantnum{}{HR0011-18-3-0007}. Any opinions, findings and conclusions or recommendations expressed in this material are those of the author(s) and do not necessarily reflect the views of the U.S. Government. Distribution Statement ``A'' (Approved for Public Release, Distribution Unlimited). We would like to acknowledge Louis Feng, Valentin Andrei, Zhongyi Lin and Serban D. Porumbescu for their feedback on early drafts of the paper.
\end{acks}

\bibliography{streamk.bib}

\clearpage
\appendix
\section{Supplementary Material}
\label{sec:appendix}

\subsection{Analytical Modeling for Stream-K Configuration}
\label{sec:analytical-modeling}
In practice, it is not always
advantageous to invoke the \emph{Stream-K} decomposition with as many CTAs
as can be actively resident on the GPU\@. Because it is a tile-splitting
approach, it incurs fixup costs above and beyond the simple
\emph{data-parallel} decomposition. Consequently, the fundamental
proposition is one of strong scaling: how much additional parallelism
can be expressed before the extra overhead causes a negative return on
investment. Depending on the problem shape, the optimal splitting could
be enough to fill the entire processor (i.e., $g \leftarrow p$), no
splitting at all (i.e., $g \leftarrow t$), or somewhere in between.

To predict this inflection point, we present a simple approach
to model the runtime of
\emph{Stream-K} as a function of grid size $g$. In the absence of
other work on the GPU, the runtime of the entire \emph{Stream-K}
schedule will be the same as that of one of its tile-outputting CTAs,
which we formulate as follows:

\begin{align*}
    time_{CTA}(g) \leftarrow & \mathpzc{a} + \mathpzc{b} (FixupPeers(g) > 1) \\
                            & + \mathpzc{c} (ItersPerCta(g)) + \mathpzc{d} (FixupPeers(g) - 1) \notag
\end{align*}

where:

\begin{align*}
    ItersPerCta(g) \leftarrow & \left\lceil
        \frac{
            \lceil \frac{m}{\text{BLK\_M}} \rceil \times
            \lceil \frac{n}{\text{BLK\_N}} \rceil \times
            \lceil \frac{k}{\text{BLK\_K}} \rceil}
        {g}\right\rceil \notag \\
    FixupPeers(g) \leftarrow &
        \left\lceil
            \frac{\left\lceil\frac{k}{\text{BLK\_K}} \right\rceil}
            {IterationsPerCta(g)}
        \right\rceil \notag
\end{align*}

This CTA runtime model comprises four components. The $\mathpzc{a}$ workload
encompasses the one-time, fixed-size costs incurred by each CTA, e.g.,
the grid launch latency, the initial compulsory cache misses, the cost
of writing the final output tile to \textbf{C}, etc. The second
component $\mathpzc{b}$ incorporates the conditional costs of outputting temporary
partial sums for scenarios where the number of output tiles does not
quantize perfectly across the processor. The third---the per-iteration
workload $\mathpzc{c}$---represents the instruction and stall workload of each
MAC-iteration. The final, per-collaborator workload $\mathpzc{d}$ is the cost of
reading and accumulating the partial sums from another CTA covering the
same tile. The set of workload constants \{$\mathpzc{a}$, $\mathpzc{b}$, $\mathpzc{c}$, $\mathpzc{d}$\} will be
unique to each combination of blocking factors, matrix data type, and
GPU microarchitecture, and can be determined empirically via
microbenchmarks.

\begin{figure}
    \centering
    \begin{subfigure}{\columnwidth}
        \includegraphics[width=\linewidth,valign=b]{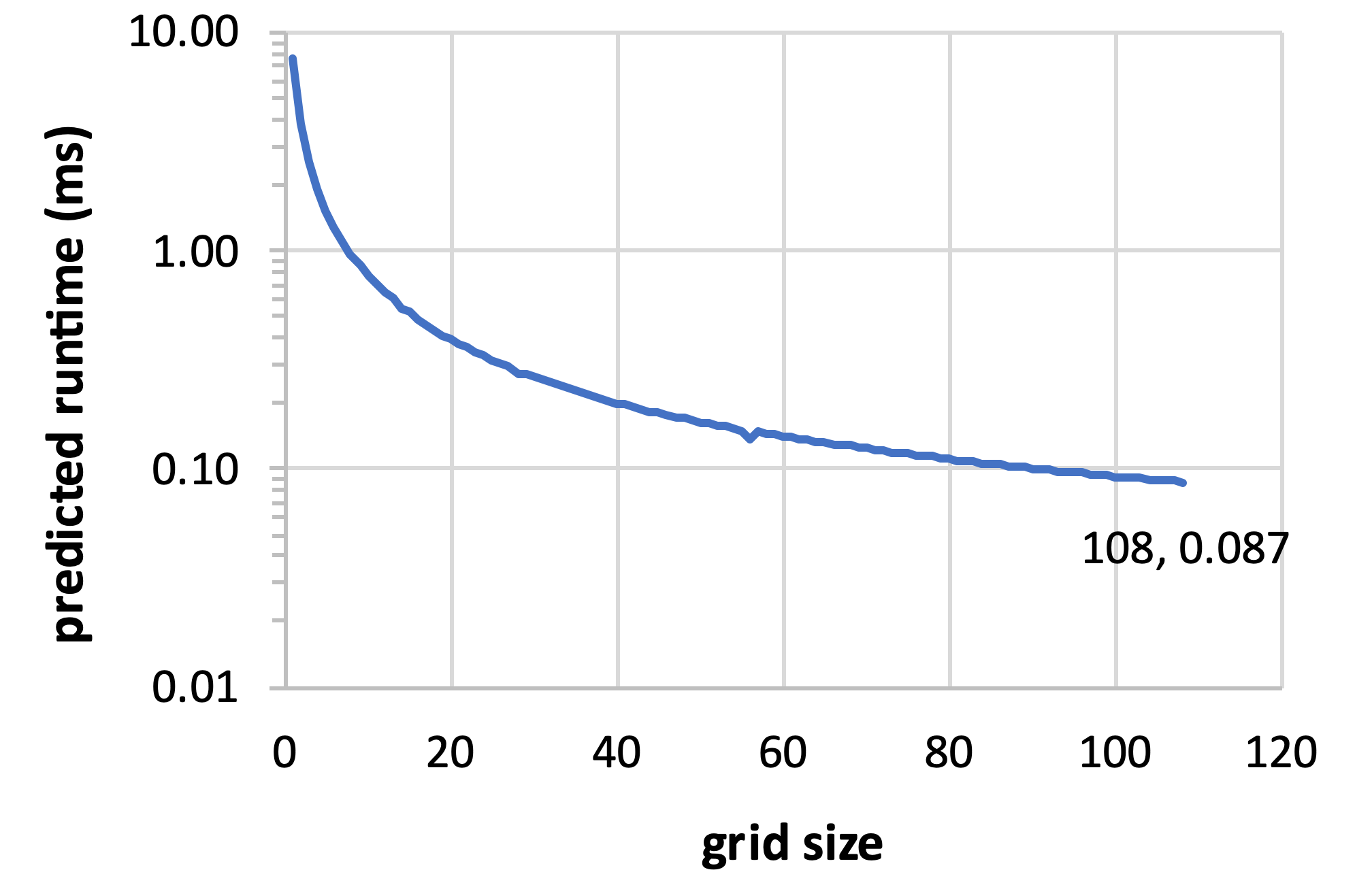}
        \caption{GEMM $256\times3584\times8192$\\
        56~output tiles, 256~iterations per tile\\
        $g_{best} \leftarrow 108$ CTAs, 132/133~iterations per CTA} \label{fig:model_smallm}
    \end{subfigure}
    \vfill%
    \begin{subfigure}{\columnwidth}
        \includegraphics[width=\linewidth,valign=b]{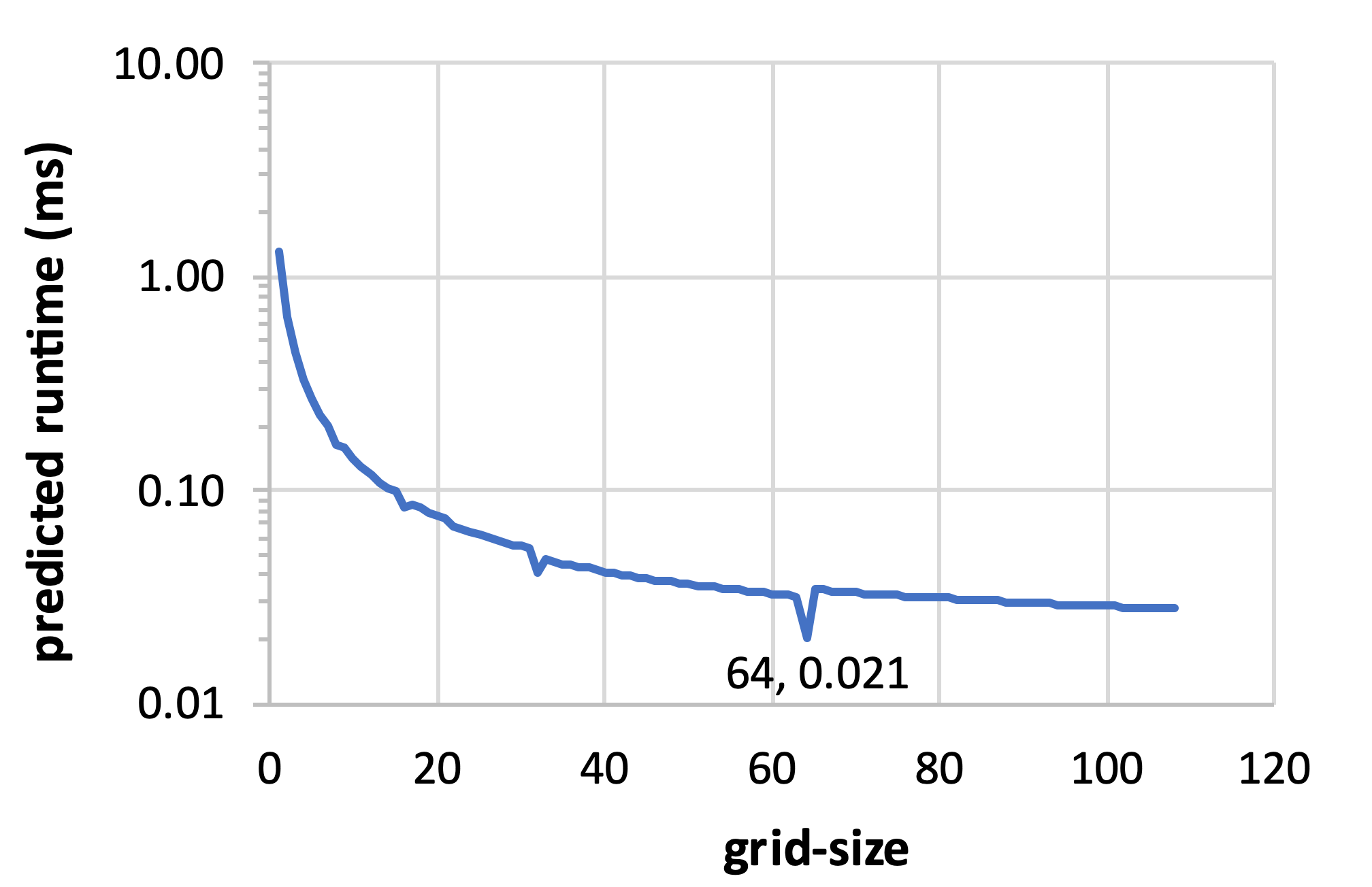}
        \caption{GEMM $1024\times1024\times1024$\\
        64~output tiles, 32~iterations per tile\\
        $g_{best} \leftarrow 64$ CTAs, 32~iterations per CTA} \label{fig:model_largemnk}
    \end{subfigure}
    \vfill%
    \begin{subfigure}{\columnwidth}
        \includegraphics[width=\linewidth,valign=b]{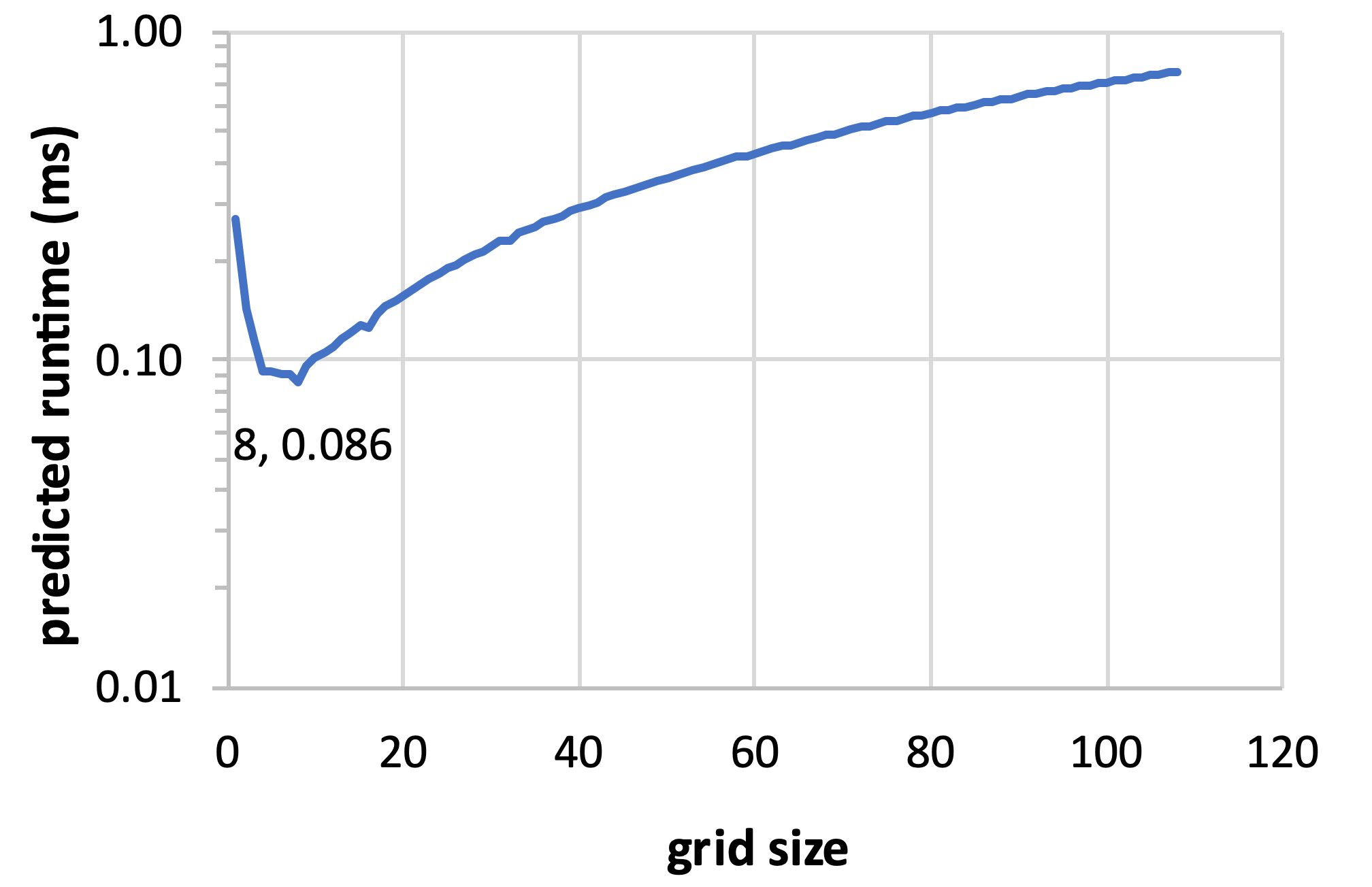}
        \caption{GEMM $128\times128\times16384$\\
        1~output tile, 512~iterations per tile\\
        $g_{best} \leftarrow 8$ CTAs, 64~iterations per CTA} \label{fig:model_largek}
    \end{subfigure}
    \caption{Modeled \textit{Stream-K} performance on NVIDIA A100 (108~SMs) for BLK\_M=128, BLK\_N=128, BLK\_K=32} \label{fig:model_streamk}
\end{figure}

Figure~\ref{fig:model_streamk} illustrates the behavior of our grid size selection model as
parameterized for fp16-precision GEMM on NVIDIA's A100 GPU using
blocking factors BLK\_M~$=128$, BLK\_N~$=128$, and BLK\_K~$=32$. Specifically, we
highlight three strong-scaling GEMM scenarios where the number of output
tiles is insufficient to produce a single full wave across the
processor's 108~SM cores.

The first GEMM shape accumulates through a large-sized
$k$-dimension to produce a short, wide output matrix. In this
scenario, the reduction in MAC-loop time relative to the increasing
costs of seam fixup is monotonically improving. Consequently, the
optimal grid size coincides with maximal parallelism at $g = 108$
CTAs.

The second shape accumulates through a medium-sized $k$-dimension
to produce a square matrix with 64 output tiles. In this case, the fixup
costs of $\mathpzc{b}$ and $\mathpzc{d}$ outweigh any reduction in MAC-loop iteration count, as
seen by the global minima ``dip'' at $g = 64$ CTAs.

\begin{figure}
    \centering
    \includegraphics[width=\columnwidth]{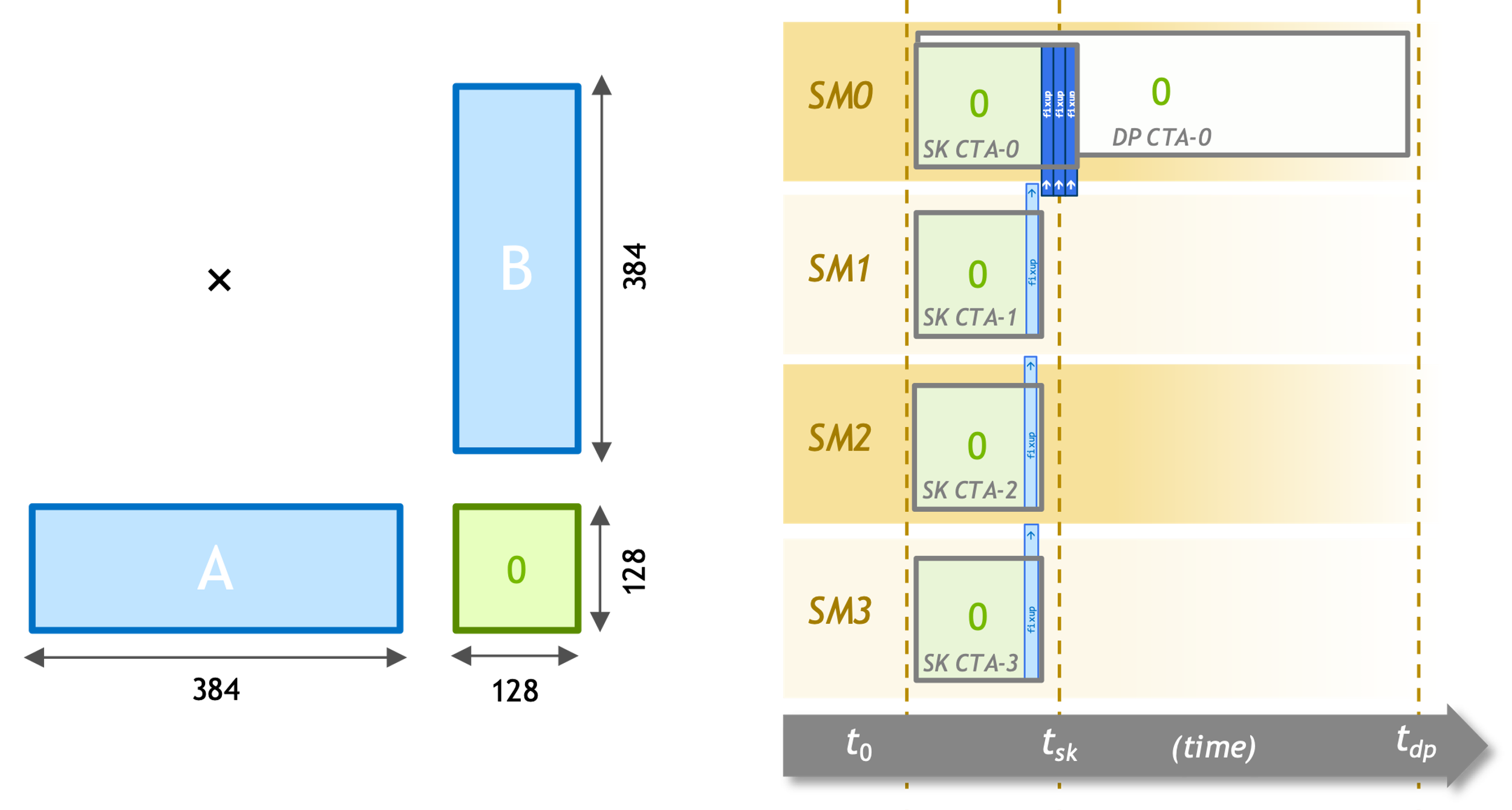}
    \caption{Strong-scaling comparison of \emph{data-parallel} and \emph{Stream-K} execution schedules for $128\times128\times384$ GEMM across a hypothetical four-SM GPU\@. \emph{Data-parallel} causes the enormous $k$-dimension to be sequentially processed within single CTA, whereas \emph{Stream-K} is able to take advantage of the parallelism available across the $k$-dimension.} \label{fig:comparison_dp_streamk}
\end{figure}

The third shape produces a single output tile after accumulating through
an enormous $k$-dimension, analogous to the execution schedule in
Figure~\ref{fig:comparison_dp_streamk}. Although the opportunity for strong scaling is quite large,
the per-peer cost of serial reduction is entirely incurred by a single
CTA\@. These accumulation costs begin to outweigh any further reductions
in iteration count for grid sizes $g>8$.



\end{document}